\documentclass[12pt]{article}
\usepackage{amsmath,amssymb}
\usepackage{bm}
\usepackage{color}
\usepackage{graphicx}
\usepackage{cite}
\usepackage{hyperref}
\usepackage{mathtools}
\usepackage{mathrsfs}
\usepackage{slashed}

\numberwithin{equation}{section}

\begin{document}

\title{
\begin{flushright}
\ \\*[-80pt]
\begin{minipage}{0.23\linewidth}
\normalsize
EPHOU-25-020\\
KYUSHU-HET-346\\*[50pt]
\end{minipage}
\end{flushright}
{\Large \bf
Modular weights of wave functions on magnetized torus
\\*[20pt]}}

\author{
Tim Jeric $^{1}$,
~Tatsuo Kobayashi $^{1}$,
~Hajime Otsuka $^{2,3}$, \\
~Maki Takeuchi $^{4}$,
and
~Hikaru Uchida $^{5}$
\\*[20pt]
\centerline{
\begin{minipage}{\linewidth}
\begin{center}
$^1${\it \normalsize
Department of Physics, Hokkaido University, Sapporo 060-0810, Japan} \\*[5pt]
$^2${\it \normalsize
Department of Physics, Kyushu University, 744 Motooka, Nishi-ku, Fukuoka 819-0395, Japan} \\*[5pt]
$^3${\it \normalsize
Quantum and Spacetime Research Institute (QuaSR), Kyushu University, 744 Motooka, Nishi-ku, Fukuoka, 819-0395, Japan} \\*[5pt]
$^4${\it \normalsize
Graduate School of Sciences and Technology for Innovation, Yamaguchi University, Yamaguchi-shi, Yamaguchi 753–8512, Japan} \\*[5pt]
$^5${\it \normalsize
National Institute of Technology, Hakodate College, Hakodate 042-0953, Japan} \\*[5pt]
\end{center}
\end{minipage}}
\\*[50pt]}

\date{
\centerline{\small \bf Abstract}
\begin{minipage}{0.9\linewidth}
\medskip
\medskip
\small
We study the origin of modular weights of wave functions in magnetized $T^{2}$ models. 
It is explicitly demonstrated that the modular weights of the wave functions on magnetized $T^2$ is equivalent to their mass level.
We further extend this result to magnetized $T^{2g}$ models. 
As a result, we construct the wave functions of excited states in magnetized $T^{2g}$ models and show that their modular weights are likewise equivalent to the corresponding mass levels.
\end{minipage}
}

\begin{titlepage}
\maketitle
\thispagestyle{empty}
\end{titlepage}

\newpage

\tableofcontents


\section{Introduction}
\label{sec:intro}

In the standard model, the origin of the flavor structure such as the flavor mixing and mass hierarchies among quarks and leptons is one of significant mysteries.
As one candidate to solve the mystery,  the modular symmetry has been attractive. It is a geometrical symmetry on a compact space such as a torus and some of its orbifolds in a higher-dimensional theory like superstring theory.
The modular symmetry induces certain non-Abelian discrete flavor symmetries such as $S_3$, $A_4$, $S_4$, and $A_5$~\cite{deAdelhartToorop:2011re}.
Indeed, a lot of modular flavor models as bottom-up approaches have been studied.
(See for early works Refs.~\cite{Feruglio:2017spp,Kobayashi:2018vbk,Penedo:2018nmg,Criado:2018thu,Kobayashi:2018wkl,Novichkov:2018ovf,Novichkov:2018nkm,deAnda:2018ecu,Okada:2018yrn,Novichkov:2018yse} and reviews Refs.~\cite{Kobayashi:2023zzc,Ding:2023htn}.)
In these models, Yukawa couplings are regarded as modular forms, which are functions of the geometrical parameter called moduli.\footnote{Several modular forms have been constructed, e.g. for $S_3$, $A_4$, $S_4$, $A_5$ and their covering groups \cite{Feruglio:2017spp,Kobayashi:2018vbk,Penedo:2018nmg,Novichkov:2018nkm,Liu:2019khw,Novichkov:2020eep,Liu:2020akv,Liu:2020msy}.}
Thus, Yukawa couplings also transform non-trivially under the modular flavor transformation.
Once the moduli are fixed, the flavor symmetry is broken and then the flavor structure can be determined.
However, several assumptions are introduced in the bottom-up models, including the choice of the modular flavor group, the representations and modular weights assigned to matter fields, and the identification of modular forms with Yukawa couplings.

As an effective field theory of magnetized D-brane model in type-IIB superstring theory, ten-dimensional ${\cal N}=1$ supersymmetric Yang-Mills theory with non-vanishing magnetic fluxes on a compact space such as torus and its orbifolds~\cite{Bachas:1995ik,Blumenhagen:2000wh,Angelantonj:2000hi,Blumenhagen:2000ea} is interesting.
It is because the magnetic fluxes generate multi-generational chiral fermions~\cite{Cremades:2004wa,Abe:2008fi,Abe:2013bca,Abe:2014noa,Kobayashi:2017dyu,Sakamoto:2020pev,Antoniadis:2009bg,Kikuchi:2022lfv,Kikuchi:2022psj,Kikuchi:2023awm,Abe:2008sx}, which can be identified with quarks and leptons, on the four-dimensional effective theory.
In addition, once we obtain wave functions of fields such as quarks, leptons, and Higgs modes on a compact space, one can derive Yukawa couplings~\cite{Cremades:2004wa,Antoniadis:2009bg,Kikuchi:2022lfv,Fujimoto:2016zjs} and higher-order couplings~\cite{Abe:2009dr} in the four-dimensional effective theory by calculating overlap integrals of the corresponding wave functions.
By using the Yukawa couplings, realistic masses, mixing angles, and CP phases of quarks and leptons can be realized as in Refs.~\cite{Fujimoto:2016zjs,Abe:2012fj,Abe:2014vza,Kobayashi:2016qag,Buchmuller:2017vho,Buchmuller:2017vut,Kikuchi:2021yog,Kikuchi:2022geu,Hoshiya:2022qvr}.
In addition, the Yukawa couplings in magnetized torus as well as some of its orbifolds become modular forms~\cite{Kobayashi:2017dyu,Kobayashi:2018rad}.
Then, the modular symmetry in magnetized torus models has been studied in detail in Refs.~\cite{
Kikuchi:2020frp,Kikuchi:2021ogn,Kikuchi:2023awe}.
(See also Refs.~\cite{Kobayashi:2016ovu,Kobayashi:2018bff,Kariyazono:2019ehj,Kobayashi:2020hoc,Ohki:2020bpo,Kikuchi:2020nxn,Almumin:2021fbk,Tatsuta:2021deu,Kikuchi:2022bkn}.)\footnote{See Ref.~\cite{Lauer:1990tm,Lerche:1989cs,Ferrara:1989qb,Baur:2019kwi,Baur:2019iai,Nilles:2020nnc,Nilles:2020kgo,Nilles:2020tdp,Kobayashi:2024ysa,Kobayashi:2024hkk} for heterotic orbifold models.}
From their analysis, it is found that zero-mode wave functions behave as modular forms of weight $1/2$ for a proper congruence subgroup, $G_d$, of the modular (sub-)group, $\Gamma$, which can be determined by the magnitude of the magnetic flux.
Then, the wave functions transform non-trivially under the finite modular (sub-)group, $\Gamma/G_d$, which becomes the modular flavor group.
Hence, the modular flavor group can be determined by the magnetic flux.
On the other hand, the origin of the modular weight has not been clarified yet, although it has been suggested that the modular weight corresponds to the mass level, as discussed in Ref.~\cite{Kikuchi:2023clx}.
In this paper, we show that the modular weight is equivalent to the mass level by extending the ladder operators related to the modular weight in Ref.~\cite{Ding:2020zxw} and comparing them with the ladder operators related to the mass level.

This paper is organized as follows.
In section~\ref{sec:T2}, we show the equivalence of the mass level and the modular weight of the wave functions in magnetized $T^2$ models.
In subsection~\ref{subsec:GeometryT2}, We review the geometry of $T^2$, shown in Ref.~\cite{Ding:2020zxw}.
In subsection~\ref{subsec:periodicmodularform}, we review periodic functions of modular forms following Ref.~\cite{Ding:2020zxw}. 
In subsection~\ref{subsec:wavefunction}, we review wave functions on magnetized $T^2$ following Refs.~\cite{Cremades:2004wa,Berasaluce-Gonzalez:2012abm,Hamada:2012wj}.
In subsection~\ref{subsec:periodicwavefunction}, we extend the analysis of periodic functions of modular forms to wave functions.
Finally, in subsection~\ref{subsec:modularweightmasslevel}, we show the equivalence of the mass level and the modular weight by comparing the ladder operators related to the mass level in subsection~\ref{subsec:wavefunction} and the ladder operators related to the modular weight in subsection~\ref{subsec:periodicwavefunction}.
Then, we extend the analysis of magnetized $T^2$ models in section~\ref{sec:T2} to magnetized $T^{2g}$ (including $T^4$ and $T^6$) models in section~\ref{sec:T2g}.
In subsection~\ref{subsec:GeometryT2g}, We review the geometry of $T^{2g}$, as shown in Ref.~\cite{Ding:2020zxw}.
In subsection~\ref{subsec:periodicSiegelmodularform}, we extend the analysis of periodic functions of modular forms shown in Ref.~\cite{Ding:2020zxw}.
In subsection~\ref{subsec:wavefunction2g}, we construct wave functions of excited states on magnetized $T^{2g}$, while wave functions of the lowest state have been studied in Refs.~\cite{Cremades:2004wa,Antoniadis:2009bg}.
In subsection~\ref{subsec:periodicwavefunctionT2g}, we extend the analysis of periodic functions of modular forms to wave functions.
Finally, in subsection~\ref{subsec:modularweightmasslevel2g}, we show the equivalence of the mass level and the modular weight by comparing the ladder operators related to the mass level in subsection~\ref{subsec:wavefunction2g} and the ladder operators related to the modular weight in subsection~\ref{subsec:periodicwavefunctionT2g}.
In section~\ref{sec:conclusion}, we conclude this paper.
In Appendix~\ref{apsec:ellEhat}, we show the detailed calculation of the equation discussed in subsection~\ref{subsec:modularweightmasslevel}.
Similarly, in Appendix~\ref{apsec:ellEhat2g}, we show the detailed calculation of the equation discussed in subsection~\ref{subsec:modularweightmasslevel2g}.
In Appendix~\ref{apsec:CFT}, we give a brief review of classical conformal field theory and compare it with modular forms and wave functions.


\section{Magnetized $T^2$ model}
\label{sec:T2}

In this section, we show that the modular weights of wave functions on a magnetized $T^2$ is equivalent to their mass level.\footnote{This possibility has already been suggested in Ref.~\cite{Kikuchi:2023clx}. In this section, we verify it explicitly.}


\subsection{Geometry of $T^2$}
\label{subsec:GeometryT2}

A two-dimensional torus, $T^2$, can be constructed by dividing the complex plane, $\mathbb{C}$ by a lattice, $\Lambda$,~i.e. $T^2 \simeq \mathbb{C}/\Lambda$.
We denote the coordinate of $\mathbb{C}$ as $u$.
The lattice can be spanned by two lattice vectors $\alpha$ and $\beta$.
Then, we denote the coordinate and the complex structure modulus of $T^2$ as $z \equiv u/\alpha$, and $\tau \equiv \beta/\alpha\ ({\rm Im}\tau > 0)$, respectively.
The identification is written as $z+1 \sim z+\tau \sim z$.
The metric of $T^2$ can be written as
\begin{align}
ds^2 = 2h_{\bar{z}^{\bar{i}}z^j}d\bar{z}^{\bar{i}}dz^{j}=|\alpha|^{2}d\bar{z}dz.
\label{eq:T2metric}
\end{align}
The Gamma matrices, $\gamma^{z}$ and $\gamma^{\bar{z}}$, are obtained from the relation, $[\gamma^{z}, \gamma^{\bar{z}}]=2h^{z\bar{z}}$, where $h^{z\bar{z}}$ is the inverse of $h_{\bar{z}z}$.
The area of $T^2$ is calculated as $A=|\alpha|^2 {\rm Im}\tau$.

An arbitrary complex structure modulus, $\tau={\rm Re}\tau+i{\rm Im}\tau=x+iy\ (y>0)$, can be generated from $\tau_0=i$ as follows.
Through the $G=SL(2,\mathbb{R})$ transformation for lattice vectors:
\begin{align}
\begin{pmatrix}
\beta \\ \alpha
\end{pmatrix}
=
\begin{pmatrix}
a & b \\
c & d
\end{pmatrix}
\begin{pmatrix}
\beta_0 \\ \alpha_0
\end{pmatrix},
\quad
g =
\begin{pmatrix}
a & b \\
c & d
\end{pmatrix}
\in G=SL(2,\mathbb{R}),
\label{eq:SL2R}
\end{align}
the complex structure modulus can be expressed as
\begin{align}
\tau = \frac{a\tau_0+b}{c\tau_0+d},
\label{eq:tau}
\end{align}
where $a$, $b$, $c$, and $d$ satisfy that $ad-bc=1$.
Note that $\tau_0=i$ is invariant under $K=SO(2) \subset G=SL(2,\mathbb{R})$.
An arbitrary $SL(2,\mathbb{R})$ matrix can be uniquely decomposed as
\begin{align}
\begin{pmatrix}
a & b \\
c & d
\end{pmatrix}
=
\begin{pmatrix}
\sqrt{y} & x/\sqrt{y} \\
0 & 1/\sqrt{y}
\end{pmatrix}
k,
\quad
y>0,
\quad
k =
\begin{pmatrix}
\cos \theta & -\sin \theta \\
\sin \theta & \cos \theta
\end{pmatrix} \in K=SO(2),
\label{eq:SL2Rdecomp}
\end{align}
and then Eq.~(\ref{eq:tau}) can be written as $\tau=x+iy$.
On the other hand, the complex coordinate, $z$, is represented from the $SL(2,\mathbb{R})$ transformation in Eq.~(\ref{eq:SL2R}) as
\begin{align}
z=\frac{z_0}{c\tau_0+d} = e^{-i\theta} y^{1/2} z_0.
\label{eq:z}
\end{align}
Hereafter, we denote $x={\rm Re}\tau$ and $y={\rm Im}\tau$.

In particular, since the lattice vectors transformed by $\Gamma=SL(2,\mathbb{Z}) \subset SL(2,\mathbb{R}) =G$ span the same lattice, $T^2$ has the $\Gamma=SL(2,\mathbb{Z})$ symmetry, which is called as the modular symmetry.
Under the $\Gamma=SL(2,\mathbb{Z})$ transformation, the complex coordinate and the complex structure modulus, $(z,\tau)$, are transformed as
\begin{align}
\gamma: (z,\tau) \rightarrow \left( \frac{z}{r \tau + s}, \frac{p \tau + q}{r \tau + s} \right), \quad
\gamma =
\begin{pmatrix}
p & q \\
r & s
\end{pmatrix}
\in \Gamma = SL(2,\mathbb{Z}),
\label{eq:modulartransformation}
\end{align}
which is called the modular transformation.


\subsection{Periodic functions of modular forms}
\label{subsec:periodicmodularform}

In this subsection, we review the analysis of section~2 in Ref.~\cite{Ding:2020zxw}.
Periodic functions, $Y(\tau, \theta)$, are defined from modular forms, $y(\tau)$, of weight $h$ for a normal subgroup of $\Gamma$, $G_d$, as
\begin{align}
Y(\tau, \theta) = j_h(g, \tau_0)^{-1} y(\tau) = e^{-ih\theta} ({{\rm Im}\tau})^{h/2} y(\tau),
\label{eq:periodY}
\end{align}
where $j_h(g, \tau_0)=(c\tau_0+d)^h=e^{ih\theta} ({{\rm Im}\tau})^{-h/2}$ denotes the automorphy factor, which satisfies the so-called cocycle condition:
\begin{align}
 j_h(g_1 g_2, \tau)= j_h(g_1, g_2\tau)  j_h(g_2, \tau).
\end{align}
They satisfy
\begin{align}
Y(\gamma'(\tau, \theta)) = Y(\tau, \theta), \ \Rightarrow \ y(\gamma' \tau) = j_h(\gamma', \tau) y(\tau), \quad \gamma' \in G_d.
\label{eq:periodicmodular}
\end{align}
In particular, classical modular forms, $y_0(\tau)$, which are holomorphic functions of $\tau$, satisfy
\begin{align}
\partial_{\bar{\tau}} y_0(\tau) = 0, \ 
\Leftrightarrow \ 
F Y_0(\tau, \theta) =0,
\label{eq:F2Y0}
\end{align}
where $F$ is defined as
\begin{align}
F
= e^{+2i\theta} \left( -2i{\rm Im}\tau \partial_{\bar{\tau}} - \frac{i}{2} \partial_{\theta} \right)
= e^{+2i\theta} \left( -2i{\rm Im}\tau \partial_{\bar{\tau}} + \frac{1}{2} e^{i\theta}\partial_{e^{i\theta}} \right).
\label{eq:F2}
\end{align}
Similarly, $E$ and $H$ are defined as
\begin{align}
E
=& e^{-2i\theta} \left( 2i{\rm Im}\tau \partial_{\bar{\tau}} + \frac{i}{2} \partial_{\theta} \right)
= e^{-2i\theta} \left( 2i{\rm Im}\tau \partial_{\bar{\tau}} - \frac{1}{2} e^{i\theta} \partial_{e^{i\theta}} \right),
\label{eq:E2} \\
H
=& i \partial_{\theta} = - e^{i\theta} \partial_{e^{i\theta}},
\label{eq:H2}
\end{align}
and they satisfy the following $SL(2,\mathbb{R})$ algebra,
\begin{align}
[ E, F ] = H, \quad [ H, E ] = + 2E, \quad [ H, F ] = - 2F.
\label{eq:SL2Ralg}
\end{align}
In addition, the periodic functions, $Y(\tau, \theta)$, satisfy
\begin{align}
H Y(\tau, \theta) = h Y(\tau, \theta).
\label{eq:eigen}
\end{align}
Therefore, the modular weight $h$ is the eigenvalue for $H$ and $E$ ($F$) raises (lowers) it by $2$.
The periodic functions of the classical modular forms, $Y_0(\tau, \theta)$, in particular, are the lowest states with respect to the modular weight.
When we write the modular weight of $Y_0(\tau, \theta)$ as $h_0$, excited states, $E^n Y_0(\tau, \theta)$, can be written as $h_0+2n$.


\subsection{Wave functions on magnetized $T^2$}
\label{subsec:wavefunction}

In this subsection, we review wave functions on magnetized $T^2$ in Refs.~\cite{Cremades:2004wa,Berasaluce-Gonzalez:2012abm,Hamada:2012wj,
Abe:2013bca,Kikuchi:2023clx}.

The $U(1)$ magnetic flux,
\begin{align}
{\cal F} = \frac{\pi iM}{{\rm Im}\tau} dz \wedge d\bar{z},
\label{eq:F}
\end{align}
is inserted on $T^2$, where the magnetic flux satisfies $(2\pi)^{-1} \int {\cal F} = M \in \mathbb{Z}$.
Hereafter, we consider $M>0$.
It is induced from the vector potential,
\begin{align}
{\cal A} = -\frac{\pi iM}{2{\rm Im}\tau}\bar{z}dz+\frac{\pi iM}{2{\rm Im}\tau}zd\bar{z}.
\label{eq:A}
\end{align}
Note that Wilson lines can be converted into Scherk-Schwarz phases~\cite{Abe:2013bca}.

Wave functions on the magnetized $T^2$ with $U(1)$ unite charge $q=1$, $\Phi$, satisfy the following boundary conditions:
\begin{align}
\Phi(z+1) &= e^{2\pi i \alpha^S}e^{\pi iM \frac{{\rm Im}z}{{\rm Im}\tau}} \Phi(z),
\label{eq:BC1} \\
\Phi(z+\tau) &= e^{2\pi i \beta^S}e^{\pi iM \frac{{\rm Im}\bar{\tau}z}{{\rm Im}\tau}} \Phi(z),
\label{eq:BCtau}
\end{align}
where $\alpha^S$ and $\beta^S$ denote the Scherk-Schwarz phases.
Under the boundary conditions, two-dimensional spinor, ${^t}(\psi_{+}, \psi_{-})$, and scalar, $\phi$, respectively satisfy the following equations of motion:
\begin{align}
i
\begin{pmatrix}
0 & 2\alpha^{-1}D_{z} \\
2\bar{\alpha}^{-1}D_{\bar{z}} & 0
\end{pmatrix}
\begin{pmatrix}
\psi_{+,n} \\ \psi_{-,n}
\end{pmatrix}
&=
m_{n,{\rm spinor}}
\begin{pmatrix}
\psi_{+,n} \\ \psi_{-,n}
\end{pmatrix},
\label{eq:eomspinor} \\
-2|\alpha|^{-2} \{ D_{z}, D_{\bar{z}} \} \phi_{n} &= m_{n,{\rm scalar}}^2 \phi_{n},
\label{eq:eomscalar}
\end{align}
where $D_{z}=\partial_{z}-i{\cal A}_{z}$ ($D_{\bar{z}}=\partial_{\bar{z}}-i{\cal A}_{\bar{z}}$) denotes the covariant derivative with $q=1$.
From Eqs.~(\ref{eq:BC1})-(\ref{eq:eomscalar}), the wave functions and mass eigenvalues can be written as
\begin{align}
\begin{pmatrix}
\psi_{+,n} \\ \psi_{-,n}
\end{pmatrix}
&=
\begin{pmatrix}
\Phi_{n} \\ \Phi_{n-1}
\end{pmatrix},
\quad
m_{n,{\rm spinor}}^2 = \frac{4\pi M}{A}n,
\label{eq:solspinor} \\
\phi_{n} &= \Phi_{n},
\qquad\quad\,\,\,
m_{n, {\rm scalar}}^2 = \frac{4\pi M}{A}\left( n+\frac{1}{2} \right),
\label{eq:solscalar}
\end{align}
where we show the explicit form of $\Phi_{n}$ in the following.
First, we define
\begin{align}
g_{+1/2} &= \frac{1}{\sqrt{2}} a = \frac{1}{\sqrt{2}} \sqrt{\frac{A}{4\pi M}} 2\bar{\alpha}^{-1} D_{\bar{z}} = e^{+i\theta} ({\rm Im}\tau)^{1/2} \frac{1}{\sqrt{2\pi M}} D_{\bar{z}},
\label{eq:g+12} \\
g_{-1/2} &= \frac{1}{\sqrt{2}} a^{\dagger} = - \frac{1}{\sqrt{2}} \sqrt{\frac{A}{4\pi M}} 2\alpha^{-1} D_{z} = e^{-i\theta} ({\rm Im}\tau)^{1/2} \frac{-1}{\sqrt{2\pi M}} D_{z},
\label{eq:g-12}
\end{align}
where $a$ and $a^{\dagger}$ satisfy $[a,a^{\dagger}]=1$, and $\theta$ denotes the argument of the lattice vector $\alpha$ on $\mathbb{C}$.
From them, one can further define
\begin{align}
2\ell_{+1} &= \{ g_{+1/2}, g_{+1/2} \} = 2\left( g_{+1/2} \right)^2,
\label{eq:ell+1} \\
2\ell_{-1} &= \{ g_{-1/2}, g_{-1/2} \} = 2\left( g_{-1/2} \right)^2,
\label{eq:ell-1} \\
2\ell_{0,{\rm scalar}} &= \{ g_{+1/2}, g_{-1/2} \} = a^{\dagger}a + \frac{1}{2},
\label{eq:ell0scalr}
\end{align}
and also
\begin{align}
f_{0} &= i
\begin{pmatrix}
0 & -g_{-1/2} \\
g_{+1/2} & 0
\end{pmatrix},
\label{eq:f0} \\
2\ell_{0, {\rm spinor}} &= \{f_0, f_{0} \} = 2\left( f_{0} \right)^2=
\begin{pmatrix}
a^{\dagger}a & 0 \\
0 & aa^{\dagger}
\end{pmatrix}.
\label{eq:ell0spinor}
\end{align}
We note that $g_{r}\ (r=\pm1/2)$ and $\ell_{n}\ (n=0, \pm1)$ in Eqs.~(\ref{eq:g+12})-(\ref{eq:ell0scalr}) satisfy the following $OSp(1|2)$ algebra~\cite{Kikuchi:2023clx}:
\begin{align}
[\ell_{m}, \ell_{n}] = (m-n) \ell_{m+n}, \quad
[\ell_{m}, g_{r}] = \left( \frac{m}{2} - r \right) g_{r}, \quad
\{ g_{r}, g_{s} \} = 2\ell_{r+s},
\label{eq:OSp12}
\end{align}
which is the subalgebra of the super-Virasoro algebra
in the Neveu-Schwarz (NS) sector, while $f_{0}$ and $\ell_{0,{\rm spinor}}$ in Eqs.~(\ref{eq:f0})-(\ref{eq:ell0spinor}) satisfy the subalgebra of the super-Virasoro algebra in the Ramond (R) sector.
In particular, the algebra of $\ell_{n}\ (n=0, \pm1)$ in the first relation in Eq.~(\ref{eq:OSp12}) is the $SL(2,\mathbb{R})$ subalgebra.
By using Eqs.~(\ref{eq:g+12})-(\ref{eq:ell0spinor}), Eqs.~(\ref{eq:eomspinor}) can be rewritten as
\begin{align}
2\ell_{0,{\rm spinor}}
\begin{pmatrix}
\psi_{+,n} \\ \psi_{-,n}
\end{pmatrix}
&= n
\begin{pmatrix}
\psi_{+,n} \\ \psi_{-,n}
\end{pmatrix},
\label{eq:eomspinorell0} \\
2\ell_{0,{\rm scalar}} \phi_{n}
&= \left( n+ \frac{1}{2} \right) \phi_{n}.
\label{eq:eomscalarell0}
\end{align}
It means that the eigenvalue for $2\ell_{0}$ is the mass level.
In addition, since the following relations,
\begin{align}
[2\ell_{0}, \ell_{\pm1}] = \mp 2 \ell_{\pm1}, \quad [2\ell_{0}, g_{\pm1/2}] = \mp g_{\pm1/2},
\label{eq:ladder}
\end{align}
are satisfied, $g_{-1/2}$ ($g_{+1/2}$) and $\ell_{-1}$ ($\ell_{+1}$) raises (lowers) the mass level by $1$ and $2$, respectively.
Therefore, the $j(\in \mathbb{Z}/M\mathbb{Z})$ th wave functions of level $n$, $\Psi^{j}_{n}$, can be written as
\begin{align}
\Phi^{j}_{n}(z,\tau, \theta)
&= \frac{1}{\sqrt{n!}} \left( \sqrt{2}g_{-1/2} \right)^n \Phi^{j}_{0}(z,\tau,\theta) \notag \\
&= e^{-i(n+1/2)\theta} ({\rm Im}\tau)^{(n+1/2)/2} \frac{1}{\sqrt{n!}} \left( \frac{-1}{\sqrt{\pi M}}D_{z} \right)^n \phi^{j}_0(z,\tau) \notag \\
&= e^{-i(n+1/2)\theta} ({\rm Im}\tau)^{(n+1/2)/2} \phi^{j}_{n}(z,\tau),
\label{eq:Phin} \\
\Phi^{j}_{0}(z,\tau,\theta)
&= e^{-i\theta/2} ({\rm Im}\tau)^{1/4} \phi^{j}_{0}(z,\tau) \notag \\
&= e^{-i\theta/2} ({\rm Im}\tau)^{1/4} \left( \frac{2M}{A^2} \right)^{1/4} e^{2\pi i\frac{j+\alpha^S}{M}\beta^S} e^{\pi iMz\frac{{\rm Im}z}{{\rm Im}\tau}} \vartheta
\begin{bmatrix}
\frac{j+\alpha^S}{M} \\ -\beta^S
\end{bmatrix}
(Mz,M\tau) \notag \\
&= e^{-i\theta/2} ({\rm Im}\tau)^{1/4} \left( \frac{2M}{A^2} \right)^{1/4} e^{2\pi i\frac{j+\alpha^S}{M}\beta^S} e^{\pi iMz\frac{{\rm Im}z}{{\rm Im}\tau}} \sum_{\ell \in \mathbb{Z}} e^{\pi iM\tau \left( \frac{j+\alpha^S}{M} + \ell \right)^2} e^{2\pi i(Mz-\beta^S)\left( \frac{j+\alpha^S}{M} + \ell \right)},
\label{eq:Phi0}
\end{align}
where $\Phi^{j}_{0}$ are the lowest state since they satisfy
\begin{align}
g_{+1/2} \Phi^{j}_{0}(z,\tau,\theta) = \ell_{+1} \Phi^{j}_0(z,\tau,\theta) = 0.
\label{eq:Phi0rel}
\end{align}
In the next subsection, let us see the behavior of these wave functions under the modular transformation.


\subsection{Periodic functions of wave functions}
\label{subsec:periodicwavefunction}

In this subsection, we extend the analysis of the periodic functions of modular forms in section 2 of Ref.~\cite{Ding:2020zxw} to those of wave functions.

From the analysis in Refs.~\cite{Kikuchi:2020frp,Kikuchi:2021ogn},
when the case with $M \in 2\mathbb{Z}$ and $(\alpha^S, \beta^S)=(0,0)$, or the case with $M \in 2\mathbb{Z}+1$ and $(\alpha^S, \beta^S)=(1/2, 1/2)$, in particular, $\phi^{j}_{n}$ in $\Phi^{j}_{n}$ in Eq.~(\ref{eq:Phin}) behave as modular forms of weight $(n+1/2)$,\footnote{Since the definition of $\phi^{j}_{n}$ is modified from one in Ref.~\cite{Kikuchi:2020frp}, the modular weight of $\phi^{j}_{n}$ is different from one in Ref.~\cite{Kikuchi:2020frp}.}
where we take the normalization of $\phi^{j}_{n}$ as
\begin{align}
\int dzd\bar{z} \left( \phi^{j}_{n} \right)^{\ast} \phi^{k}_{n} = ({\rm Im}\tau)^{-(n+1/2)} \delta_{j,k}.
\label{eq:normalizationphi}
\end{align}
As in Eq.~(\ref{eq:periodicmodular}), $\Phi^{j}_{n}$ become the periodic functions,~i.e.,
\begin{align}
\Phi_n^{j}(\gamma'(z, \tau, \theta)) = \Phi_n^{j}(z, \tau, \theta) \quad \gamma' \in G_d,
\label{eq:periodicwave}
\end{align}
and they are canonically normalized as
\begin{align}
\int dzd\bar{z} \left( \Phi^{j}_{n} \right)^{\ast} \Phi^{k}_{n} = \delta_{j,k}.
\label{eq:normalizationPhi}
\end{align}
The lowest state, $\Phi^{j}_{0}$, also satisfies
\begin{align}
\hat{F} \Phi^{j}_{0}(z,\tau,\theta) = 0,
\label{eq:Fhat2Phi0}
\end{align}
where $\hat{F}$ is defined by modifying $F$ in Eq.~(\ref{eq:F2}) as
\begin{align}
\hat{F} 
= e^{+2i\theta} \left( -2i{\rm Im}z \partial_{\bar{z}} -2i {\rm Im}\tau \partial_{\bar{\tau}} - \frac{i}{2} \partial_{\theta} \right) 
= e^{+2i\theta} \left( -2i{\rm Im}z \partial_{\bar{z}} -2i{\rm Im}\tau \partial_{\bar{\tau}} + \frac{1}{2} e^{i\theta}\partial_{e^{i\theta}} \right).
\label{eq:Fhat2}
\end{align}
Similarly, we define $\hat{E}$ by modifying $E$ in Eq.~(\ref{eq:E2}) as
\begin{align}
\hat{E} 
= e^{-2i\theta} \left( 2i{\rm Im}z \partial_{z} +2i {\rm Im}\tau \partial_{\tau} + \frac{i}{2} \partial_{\theta} \right) 
= e^{-2i\theta} \left( 2i{\rm Im}z \partial_{z}  + 2i{\rm Im}\tau \partial_{\bar{\tau}} - \frac{1}{2} e^{i\theta} \partial_{e^{i\theta}} \right).
\label{eq:Ehat2}
\end{align}
They satisfy the following $SL(2,\mathbb{R})$ algebra:
\begin{align}
[\hat{E}, \hat{F}] = \hat{H}, \quad [\hat{H}, \hat{E}] = +2\hat{E}, \quad [\hat{H}, \hat{F}] = -2\hat{F}, \label{eq:SL2Ralghat}
\end{align}
where $\hat{H}$ is the same as $H$,
\begin{align}
\hat{H} = H = i \partial_{\theta} = - e^{i\theta} \partial_{e^{i\theta}}.
\label{eq:Hhat2}
\end{align}
In addition, since $\Phi^{j}_{n}$ satisfies
\begin{align}
\hat{H} \Phi^{j}_{n}(z,\tau,\theta) = \left( n+\frac{1}{2} \right) \Phi^{j}_{n}(z,\tau,\theta),
\label{eq:Hhat2Phi}
\end{align}
we can indeed obtain the modular weight of $\Phi^{j}_{n}$, $h=n+1/2$, as the eigenvalue of $\hat{H}$, and then it is raised (lowered) by 2 by operating $\hat{E}$ ($\hat{F}$).

We note that it is the same as the mass level of the scalar, $\phi_{n}=\Phi_n$, in Eq.~(\ref{eq:eomscalarell0}).
On the other hand, for the spinor, $(\psi_{+,n},\psi_{-,n})=(\Phi_{n},\Phi_{n-1})$, we should modify the definition of the periodic functions of the wave functions.
Recall that $z$ depends on $\theta$ in Eq.~(\ref{eq:z}).
Here, we denote $z'$ when $\theta'=\theta+\Delta\theta$.
Then, $z'$ is related to $z$ as
\begin{align}
z' = e^{-i\theta'} ({\rm Im}\tau)^{1/2} z_0 = e^{-i(\theta+\Delta\theta)} ({\rm Im}\tau)^{1/2} z_0 = e^{-i\Delta\theta} z.
\label{zprimez}
\end{align}
Now, let us reconsider the definition of periodic functions of wave functions.
In the case of scalar $\phi$, the periodic function, $\Phi$, with the modular weight $h$ satisfies that
\begin{align}
\Phi(z',\tau,\theta')
=& e^{-ih\theta'} ({\rm Im}\tau)^{h/2} \phi(z',\tau) \notag \\
=& e^{-ih(\theta+\Delta\theta))} ({\rm Im}\tau)^{h/2} \phi(e^{-i\Delta\theta}z,\tau) \notag \\
=& e^{-ih\Delta\theta} \left( e^{-ih\theta} ({\rm Im}\tau)^{h/2} \phi(z,\tau) \right) \notag \\
=& e^{-ih\Delta\theta} \Phi(z,\tau,\theta),
\label{eq:rePhi}
\end{align}
where the scalar satisfies that
\begin{align}
\phi(e^{-i\Delta\theta}z,\tau) = \phi(z,\tau).
\label{eq:scalar}
\end{align}
In the case of a spinor $\psi_{\pm}$, on the other hand, the periodic function, $\Psi_{\pm}$ with the modular weight $h_{\pm}$ satisfies that
\begin{align}
\Psi_{\pm}(z',\tau,\theta')
=& e^{-ih_{\pm}\theta'} ({\rm Im}\tau)^{h_{\pm}/2} \psi_{\pm}(z',\tau) \notag \\
=& e^{-ih_{\pm}(\theta+\Delta\theta))} ({\rm Im}\tau)^{h_{\pm}/2} \psi_{\pm}(e^{-i\Delta\theta}z,\tau) \notag \\
=& e^{-ih_{\pm}\Delta\theta} e^{\mp i\Delta\theta/2}\left( e^{-ih_{\pm}\theta} ({\rm Im}\tau)^{h_{\pm}/2} \psi_{\pm}(z,\tau) \right) \notag \\
=& e^{-i\left(h_{\pm} \pm 1/2\right)\Delta\theta} \Psi_{\pm}(z,\tau,\theta),
\label{eq:rePsi}
\end{align}
where the spinor satisfies that
\begin{align}
\psi_{\pm}(e^{-i\Delta\theta}z,\tau) = e^{\mp i\Delta\theta/2} \psi_{\pm}(z,\tau).
\label{eq:spinor}
\end{align}
From Eqs.~(\ref{eq:rePhi}) and (\ref{eq:rePsi}),
when the periodic function of the wave function of the spinor, $\Psi_{\pm}$, is the same as the wave function of the scalar, $\Phi$, the modular weight of the spinor, $h_{\pm}$, can be written by the modular weight of the scalar, $h$, as $h_{\pm}=h \mp 1/2$.
In this case, the spinor, $\psi_{\pm}$, in Eq.~(\ref{eq:rePsi}), which satisfies Eq.~(\ref{eq:scalar}), can be represented by the scalar, $\phi$, in Eq.~(\ref{eq:rePhi}), which satisfies Eq.~(\ref{eq:spinor}), as
\begin{align}
\psi_{\pm}(z,\tau) = e^{\mp i\theta/2} \phi(z,\tau).
\label{eq:spinorscalar}
\end{align}
Hence, the modular weight of the spinor, $(\psi_{+,n}, \psi_{-,n})=(\Phi_{n}, \Phi_{n-1})$, can be obtained as
\begin{align}
\hat{H}
\begin{pmatrix}
\psi_{+,n} \\ \psi_{-,n}
\end{pmatrix}
=
\begin{pmatrix}
h_{+} & 0 \\
0 & h_{-}
\end{pmatrix}
\begin{pmatrix}
\psi_{+,n} \\ \psi_{-,n}
\end{pmatrix}
=
\begin{pmatrix}
\left(n+\frac{1}{2}\right)-\frac{1}{2} & 0 \\
0 & \left((n-1)+\frac{1}{2}\right)+\frac{1}{2}
\end{pmatrix}
\begin{pmatrix}
\psi_{+,n} \\ \psi_{-,n}
\end{pmatrix}
= n
\begin{pmatrix}
\psi_{+,n} \\ \psi_{-,n}
\end{pmatrix}.
\label{eq:Hhat2spinor}
\end{align}
We note that it is also the same as the mass level of the spinor in Eq.~(\ref{eq:eomspinorell0}).

Finally, we give a brief comment from the viewpoint of two-dimensional conformal field theory. (See Appendix \ref{apsec:CFT}.) The behavior of the periodic function $\Phi$ in Eq.~(\ref{eq:rePhi}) is similar to Eq.~(\ref{eq:CFT}) in two-dimensional conformal field theory. From such a viewpoint, Eq.~(\ref{eq:spinorscalar}) may correspond to the relation between the NS and R sectors. Furthermore, excited states may correspond to oscillating modes in string theory, where the oscillating operator $\partial X$ increases conformal dimension by one. We would study these issues more, including twisted boundary conditions on orbifolds elsewhere.


\subsection{Modular weight versus Mass level}
\label{subsec:modularweightmasslevel}

In this subsection, we show that the modular weight of the wave functions on a magnetized $T^2$ is equivalent to their mass level.
Here, we consider the scalar, $\phi_{n}=\Phi_{n}$.

From Eqs.~(\ref{eq:Hhat2Phi}) and (\ref{eq:eomscalarell0}), both the modular weight (which is the eigenvalue of $\hat{H}$) and the mass level (which is the eigenvalue of $2\ell_{0,{\rm scalar}}$) are given by $n+1/2$.
From Eqs.~(\ref{eq:SL2Ralghat}) and (\ref{eq:ladder}), the modular weight and the mass level are raised (lowered) by 2 by operating $\hat{E}$ ($\hat{F}$) and $\ell_{-1}$ ($\ell_{+1}$), respectively.
In particular, from Eqs.~(\ref{eq:Fhat2Phi0}) and (\ref{eq:Phi0rel}), $\Phi_{0}$ becomes the lowest state with the meaning of both the modular weight and the mass level.
Now, as shown in Eq.~(\ref{eq:ellEhatdetail}) of Appendix~\ref{apsec:ellEhat} in detail, we can show that operating $\ell_{-1}$ on $\Phi_{0}$ is equivalent to operating $\hat{E}$ on $\Phi_{0}$,
\begin{align}
\ell_{-1} \Phi^j_0(z,\tau,\theta) = \hat{E} \Phi^j_0(z,\tau,\theta).
\label{eq:ellEhat}
\end{align}
Similarly, operating $\ell_{+1}$ is equivalent to $-\hat{F}$.\footnote{Indeed, it is consistent that $[\ell_{+1}, \ell_{-1}]=2\ell_{0,{\rm scalar}}$ and $[-\hat{F}, \hat{E}]=\hat{H}$.}
In other words, $\ell_{-1}$ ($\ell_{+1}$) raises (lowers) not only the mass level but the modular weight by 2.
Furthermore, $g_{-1/2}$ ($g_{+1/2}$) also raises (lowers) not only the mass level but the modular weight by 1.
Indeed, it is satisfied that
\begin{align}
[\hat{H}, \ell_{\pm1}] = \mp 2 \ell_{\pm1}, \quad [\hat{H}, g_{\pm1/2}] = \mp g_{\pm1/2}.
\label{eq:Hhat2ellg}
\end{align}
Therefore, it is found that the modular weight of the wave functions on magnetized $T^2$ is equivalent to their mass level.


\section{Magnetized $T^{2g}$ model}
\label{sec:T2g}

In this section, we extend the analysis of a magnetized $T^2$ model in the previous section to a magnetized $T^{2g}$ model.
For that purpose, we further extend the analysis of Ref.~\cite{Ding:2020zxw}  in subsection~\ref{subsec:periodicSiegelmodularform},
and we study wave functions of the excited states on the magnetized $T^{2g}$ in subsection~\ref{subsec:wavefunction2g}.


\subsection{Geometry of $T^{2g}$}
\label{subsec:GeometryT2g}

A $2g$-dimensional torus, $T^{2g}$, can be constructed by dividing $\mathbb{C}^{g}$ by a lattice, $\Lambda$,~i.e. $T^{2g} \simeq \mathbb{C}^{g}/\Lambda$.
We denote the coordinates of $\mathbb{C}^{g}$ as $u^{I}\ (I=1,...,g)$.
The lattice can be spanned by $2g$ lattice vectors $\alpha_i$ and $\beta_i$ $(i=1,...,g)$, whose coordinates are written as $\alpha_i={^t}(\alpha_{i1},...,\alpha_{ig})$ and $\beta_{i}={^t}(\beta_{i1},...,\beta_{ig})$, respectively.
Then, we denote the coordinates and the complex structure moduli of $T^{2g}$ as $z^{i} \equiv (\alpha^{-1})^{i}_{I}u^{I}$, and $\Omega_{ij} \equiv (\alpha^{-1}\beta)_{ij}\ ({\rm Im}\Omega > 0,\ {^t}\Omega=\Omega)$, respectively.
The identification is written as $z+e_k \sim z+\Omega e_k \sim z$.
The metric of $T^2$ can be written as
\begin{align}
ds^2 = 2h_{z^{i}\bar{z}^{\bar{j}}}dz^{i}d\bar{z}^{\bar{j}}={^t}(\alpha^{\dagger}\alpha)_{i\bar{j}}dz^{i}d\bar{z}^{\bar{j}}.
\label{eq:T2gmetric}
\end{align}
The Gamma matrices, $\Gamma^{z^{i}}$ and $\Gamma^{\bar{z}^j}$, are obtained from the relation, $[\Gamma^{z^{i}}, \Gamma^{\bar{z}^{j}}]=2h^{z^{i}\bar{z}^{\bar{j}}}$, where $h^{z^{i}\bar{z}^{\bar{j}}}$ is the inverse matrix of $h_{\bar{z}^{\bar{i}}z^{j}}$.
The volume of $T^{2g}$ is $V=|{\rm det}(\alpha^{\dagger}\alpha)| {\rm det}({\rm Im}\Omega)$.

Arbitrary complex structure moduli, $\Omega=X+iY$, can be generated from $\Omega_0=\alpha_0^{-1}\beta_{0}=iI_{g}$ as follows.
Through the $G=Sp(2g,\mathbb{R})$ transformation for the lattice vectors:
\begin{align}
\begin{pmatrix}
{^t}\beta \\ {^t}\alpha
\end{pmatrix}
=
\begin{pmatrix}
A & B \\
C & D
\end{pmatrix}
\begin{pmatrix}
{^t}\beta_0 \\ {^t}\alpha_0
\end{pmatrix},
\quad
g =
\begin{pmatrix}
A & B \\
C & D
\end{pmatrix}
\in G=Sp(2g,\mathbb{R}),
\label{eq:Sp2gR}
\end{align}
the complex structure moduli can be expressed as
\begin{align}
\Omega = (A\Omega_0+B)(C\Omega_0+D)^{-1},
\label{eq:Omega}
\end{align}
where $A$, $B$, $C$, and $D$ satisfy that
\begin{align}
A{^t}C=C{^t}A, \quad B{^t}D=D^{^t}B, \quad A{^t}D-C{^t}B=I. \label{eq:Sp2gRcond}
\end{align}
Note that $\Omega_0=iI_{g}$ is invariant under $K=Sp(2g,\mathbb{R}) \cap O(2g,\mathbb{R}) \simeq U(g)$.
In other words, there are $K \simeq U(g)$ degrees of freedom in the choice of $\alpha_0$.~\footnote{In subsection~\ref{subsec:wavefunction2g}, we choose $\alpha_0$ such that they satisfy $({O'}^{-1}\alpha_0^{\dagger}\alpha_0{O'})_{\bar{i}j}=|\alpha_0|_{j}^2 \delta_{\bar{i},j}$, where $O'$ is an orthogonal matrix defined later and the eigenvalue satisfies that $V=\prod_{k=1}^{g} |\alpha_0|_{k}^2$.}
An arbitrary $Sp(2g,\mathbb{R})$ matrix can be uniquely decomposed as
\begin{align}
\begin{pmatrix}
A & B \\
C & D
\end{pmatrix}
=
\begin{pmatrix}
\sqrt{Y} & X\sqrt{Y}^{-1} \\
0 & \sqrt{Y}^{-1}
\end{pmatrix}
k,
\quad
k =
\begin{pmatrix}
A' & -B' \\
B' & A'
\end{pmatrix} \in K=Sp(2g,\mathbb{R}) \cap O(2g,\mathbb{R}),
\label{eq:Sp2gRdecomp}
\end{align} 
and then Eq.~(\ref{eq:Omega}) can be written as $\Omega=X+iY$.
Here, since $X$ and $Y$ commute with each other due to Eq.~(\ref{eq:Sp2gRcond}), $X$, $Y$, and $\Omega$ are simultaneously diagonalized by a real orthogonal matrix, $O$,~i.e.,
\begin{align}
\Omega =& X + i Y, \notag \\
\Rightarrow \ 
O {\rm diag}(\tau)O^{-1} =& O {\rm diag}(x) O^{-1} + i O {\rm diag}(y) O^{-1},
\label{eq:Omegadiagonalize}
\end{align}
where $\tau_j=x_j + i y_j = {\rm Re}\tau_j + i{\rm Im}\tau_j$ $(j=1,...,g)$ are the eigenvalues.
In addition, we note that $\Omega$ is also invariant under $K$.
It means that there are $K \simeq U(g)$ degrees of freedom in the choice of $\alpha$.
The complex coordinate, $z$, on the other hand, is represented from the $G=Sp(2g,\mathbb{R})$ transformation for the coordinate, $z_0=\alpha_0^{-1}u$, as
\begin{align}
z=\alpha^{-1}u={^t}(C\Omega_0+D)^{-1}\alpha_0^{-1}u = \sqrt{Y} {^t}U^{-1} z_0, \quad U=A'+iB' \in U(g).
\label{eq:z2g}
\end{align}
Hereafter, we denote $X={\rm Re}\Omega$ and $Y={\rm Im}\Omega$.

In particular, since the lattice vectors transformed by $\Gamma_{g}=Sp(2g,\mathbb{Z}) \subset Sp(2g,\mathbb{R}) =G$ span the same lattice, $T^{2g}$ has the $\Gamma_{g}=Sp(2g,\mathbb{Z})$ symmetry, which is called as the Siegel modular symmetry.
Under the $\Gamma_{g}=Sp(2g,\mathbb{Z})$ transformation, the complex coordinate and the complex structure moduli, $(z,\Omega)$, are transformed as
\begin{align}
\gamma: (z,\Omega) \rightarrow \left( {^t}(R \Omega + S)^{-1}z, (P \Omega + Q)(R \Omega + S)^{-1} \right), \quad
\gamma =
\begin{pmatrix}
P & Q \\
R & S
\end{pmatrix}
\in \Gamma_{g} = Sp(2g,\mathbb{Z}),
\label{eq:modulartransformation2g}
\end{align}
which is called the Siegel modular transformation.



\subsection{Periodic functions of Siegel modular forms}
\label{subsec:periodicSiegelmodularform}

In this subsection, we further extend the analysis of Ref.~\cite{Ding:2020zxw}.
Periodic functions, $Y(\Omega, U)$, are defined from Siegel modular forms, $y(\Omega)$, of weight $h$ for a normal subgroup of $\Gamma_{g}$, $G_d$, as
\begin{align}
Y(\Omega, U) = j_h(g, \Omega_0)^{-1} y(\Omega) = [{\rm det}(U)]^{-h} [{\rm det}({{\rm Im}\Omega})]^{h/2} y(\Omega),
\label{eq:periodY2g}
\end{align}
where we take the automorphy factor as $j_h(g, \Omega_0)=[{\rm det}(C\Omega_0+D)]^h=[{\rm det}(U)]^h [{\rm det}({{\rm Im}\Omega})]^{-h/2}$.
They satisfy
\begin{align}
Y(\gamma'(\Omega, U)) = Y(\Omega, U), \ \Rightarrow \ y(\gamma' \Omega) = j_h(\gamma', \Omega) y(\Omega), \quad \gamma' \in G_d.
\label{eq:periodicSiegelmodular}
\end{align}
In particular, classical Siegel modular forms, $y_0(\Omega)$, which are holomorphic functions of $\Omega$, satisfy
\begin{align}
\partial_{\bar{\tau}_j} y_0(\Omega) = 0, \ 
\Leftrightarrow \ 
F_j Y_0(\Omega, U) =0,
\ (\forall j=1,...,g)
\label{eq:F2gY0}
\end{align}
where $F_j$ is defined as
\begin{align}
F_j
= e^{+2i\theta_j} \left( -2i{\rm Im}\tau_j \partial_{\bar{\tau}_j} + \frac{1}{2} e^{i\theta_j}\partial_{e^{i\theta_j}} \right),
\label{eq:Fj2g}
\end{align}
where we do not sum over $j$.
Here, $\tau_j$ and $e^{i\theta_j}$ denote the eigenvalues of $\Omega$ and $U$, respectively.
Similarly, $E_j$ and $H_j$ are defined as
\begin{align}
E_j
&= e^{-2i\theta_j} \left( 2i{\rm Im}\tau_j \partial_{\bar{\tau}_j} - \frac{1}{2} e^{i\theta_j} \partial_{e^{i\theta_j}} \right),
\label{eq:Ej2g} \\
H_j
&= - e^{i\theta_j} \partial_{e^{i\theta_j}},
\label{eq:H2g}
\end{align}
and they satisfy the following $SL(2,\mathbb{R})_j$ algebra:
\begin{align}
[ E_j, F_j ] = H_j, \quad [ H_j, E_j ] = + 2E_j, \quad [ H_j, F_j ] = - 2F_j.
\label{eq:SL2Ralg2g}
\end{align}
It means that $E_j$ ($F_j$) raises (lowers) the eigenvalue of $H_j$ by $2$.
In addition, the periodic functions, $Y(\Omega, U)$, satisfy
\begin{align}
H_j Y(\Omega, U) = h Y(\Omega, U).
\label{eq:eigen2g}
\end{align}
Here, when we consider the automorphy factor as $j_h(g, \Omega_0)=[{\rm det}(C\Omega_0+D)]^h$, 
any eigenvalues of $H_j$ for $j=1,...,g$ are the same as the modular weight $h$.
Indeed, the periodic functions of the classical Siegel modular forms, $Y_0(\Omega, U)$ are the lowest states with respect to the modular weight, and the eigenvalues of $H_j$ for $j=1,...,g$ are the same as the modular weight $h_0$.
Then, to obtain $j$-invariant modular weight, $h$, as the eigenvalue of $H$ operator, we define $H$ as
\begin{align}
H = - {\rm det}U\partial_{{\rm det}U} = \frac{1}{g} \sum_{k=1}^{g} H_j.
\label{eq:Have2g}
\end{align}
In addition, to obtain the excited states such that they
behave as the Siegel modular forms with the automorphy factor $j_h(g, \Omega_0)=[{\rm det}(C\Omega_0+D)]^h$,
we define the raising operator:
\begin{align}
E = \prod_{k=1}^{g} E_k.
\label{eq:detE}
\end{align}
We also define the lowering operator:
\begin{align}
F = \prod_{k=1}^{g} F_k.
\label{eq:detF}
\end{align}
They satisfy the following conditions:
\begin{align}
[H, E] = +2 E, \quad [H, F] = -2 F,
\label{eq:HdetEFcond}
\end{align}
which means that $E$ ($F$) raises (lowers) the modular weight by $2$.
Then, the excited states are constructed by multiplying $E$ to the lowest states $Y_0(\Omega, U)$,~i.e., $E^n Y_0(\Omega, U)$, whose modular weight can be written by $h=h_0+2n$.


\subsection{Wave functions on magnetized $T^{2g}$}
\label{subsec:wavefunction2g}

In this subsection, we study wave functions of the excited states on the magnetized $T^{2g}$.\footnote{Wave functions of the lowest states have already been studied in Refs.~\cite{Cremades:2004wa,Antoniadis:2009bg}.}

The $U(1)$ magnetic flux,
\begin{align}
{\cal F} = \pi [ {^t}N ({\rm Im}\Omega)^{-1} ]_{i\bar{j}} (i dz^i \land d\bar{z}^{\bar{j}}),
\label{eq:F2g}
\end{align}
is inserted on $T^{2g}$, where it is quantized, $N_{ij} \in \mathbb{Z}$.
We also consider that the F-flat condition,
\begin{align}
{^t}(N\Omega)=N\Omega,
\label{eq:Fflat}
\end{align}
is satisfied.
In this case, the matrix, $-i{\cal F}_{z\bar{z}}$, becomes a real symmetric matrix.
When we write Eq.~(\ref{eq:F2g}) by $z_0$ in Eq.~(\ref{eq:z2g}) and $\Omega_0=iI_{g}$, it can be written as
\begin{align}
{\cal F} = \pi [ U^{-1} \sqrt{{\rm Im}\Omega} {^t}N \sqrt{{\rm Im}\Omega}^{-1} U ]_{i\bar{j}} (i dz_0^i \land d\bar{z}_0^{\bar{j}})
\equiv \pi [N_0]_{i\bar{j}} (i dz_0^i \land d\bar{z}_0^{\bar{j}}),
\label{eq:F2gz0}
\end{align}
with
\begin{align}
{^t}N_0 = N_0,
\label{eq:N0}
\end{align}
due to Eq.~(\ref{eq:Fflat}).
Here, we can rewrite the real symmetric matrix, $\sqrt{{\rm Im}\Omega} {^t}N \sqrt{{\rm Im}\Omega}^{-1}$ as follows:
\begin{align}
\sqrt{{\rm Im}\Omega} {^t}N \sqrt{{\rm Im}\Omega}^{-1}
= \sqrt{{\rm Im}\Omega {^t}N ({\rm Im}\Omega)^{-1} N}
= N.
\label{eq:N}
\end{align}
Since all of $N$, ${\rm Im}\Omega$, and $N{\rm Im}\Omega$ are real symmetric matrices, they are simultaneously diagonalized by a real orthogonal matrix, $O$.
Then, we can decompose $U$ as
\begin{align}
U = O {\rm diag}(e^{i\theta}) {O'}^{-1},
\label{eq:U}
\end{align}
and $N_0$ can be written as
\begin{align}
N_0 = U^{-1} N U = {O'} {\rm diag}(M) {O'}^{-1}.
\label{eq:N0re}
\end{align}

The magnetic flux
in Eq.~(\ref{eq:F2g}) 
is induced from the vector potential:
\begin{align}
{\cal A} = - \frac{\pi i}{2}[ {^t}\bar{z}{^t}N ({\rm Im}\Omega)^{-1} ]_k dz^k + \frac{\pi i}{2} [ {^t}z{^t}N ({\rm Im}\Omega)^{-1} ]_k d\bar{z}^{\bar{k}}.
\label{eq:A2g}
\end{align}
Note that Wilson lines can be converted into Scherk-Schwarz phases~\cite{Abe:2013bca}.
When we write it by $z_0$ in Eq.~(\ref{eq:z2g}) and $\Omega_0=iI_{g}$, it can be written as
\begin{align}
{\cal A} = - \frac{\pi i}{2}[ {^t}\bar{z}_0 N_0 ]_k dz_0^k + \frac{\pi i}{2} [ {^t}z_0 N_0 ]_k d\bar{z}_0^{\bar{k}}.
\label{eq:A2gz0}
\end{align}

The covariant derivatives with $U(1)$ charge $q$, $D_{z} = \partial_{z}-iq{\cal A}_{z}$ and $D_{\bar{z}}=\partial_{\bar{z}}-iq{\cal A}_{\bar{z}}$, satisfy
\begin{align}
&[D_{z^{i}}, D_{\bar{z}^{\bar{j}}}] = -iq{\cal F}_{z^{i}\bar{z}^{\
bar{j}}} = \pi q [{^t}N({\rm Im}\Omega)^{-1}]_{i\bar{j}},
\label{eq:DDbarF} \\
\Rightarrow \ 
&[(2 {^t}\alpha^{-1} {^t}D_{z})_{I}, (2 D_{\bar{z}} {^t}(\alpha^{\dagger})^{-1})_{\bar{J}}] = 4\pi q [{^t}\alpha^{-1}{^t}N({\rm Im}\Omega)^{-1}{^t}(\alpha^{\dagger})^{-1}]_{I\bar{J}},
\notag \\
\Rightarrow \ 
&[(2 {^t}\alpha_0^{-1} {^t}D_{z_0})_{I}, (2 D_{\bar{z}_0} {^t}(\alpha_0^{\dagger})^{-1} )_{\bar{J}}] = 4\pi q [{^t}\alpha_0^{-1} {O'} {\rm diag}(M) {O'}^{-1} {^t}(\alpha_0^{\dagger})^{-1}]_{I\bar{J}}.
\label{eq:DDbarFre}
\end{align}
Here, we choose $\alpha_0$ such that it satisfies that
\begin{align}
{O'}^{-1} {^t}(\alpha_0^{\dagger} \alpha_0)^{-1} {O'} = {\rm diag}(|\alpha_0|^{-2}).
\label{eq:alpha0}
\end{align}
In this case, Eq.~(\ref{eq:DDbarFre}) is further rewritten as
\begin{align}
&\left[ \left(\frac{2}{\sqrt{4\pi q}}{\rm diag}\left(\sqrt{M^{-1}}\right) {O'}^{-1} {^t}D_{z_0} \right)_{I'}, \left( \frac{2}{\sqrt{4\pi q}} D_{\bar{z}_0} {O'} {\rm diag}\left(\sqrt{M^{-1}}\right) \right)_{J'} \right] = \delta_{I',J'}.
\label{eq:aadagger}
\end{align}
Hence, when we define
\begin{align}
g_{+1/2}
&= \frac{1}{\sqrt{2}} a \notag \\
&= \frac{2}{\sqrt{8\pi q}} D_{\bar{z}_0} {O'} {\rm diag}\left(\sqrt{M^{-1}}\right) \notag \\
&= \frac{1}{\sqrt{2\pi q}} D_{\bar{z}} O {\rm diag}\left(\sqrt{M^{-1}{\rm Im}\tau} e^{i\theta} \right) \label{eq:g+122g} \\
&= \frac{1}{\sqrt{2\pi q}} D_{\bar{z}} \sqrt{ N^{-1} {\rm Im}\Omega} U {O'}, \notag
\end{align}
\begin{align}
g_{-1/2}
&= \frac{1}{\sqrt{2}} a^{\dagger} \notag \\
&= -\frac{2}{\sqrt{8\pi q}} D_{z_0} {O'} {\rm diag}\left(\sqrt{M^{-1}}\right) \notag \\
&= -\frac{1}{\sqrt{2\pi q}} D_{z} O {\rm diag}\left( \sqrt{M^{-1} {\rm Im}\tau} e^{-i\theta} \right) \label{eq:g-122g} \\
&= -\frac{1}{\sqrt{2\pi q}} D_{z} \sqrt{N^{-1} {\rm Im}\Omega} {^t}U^{-1} {O'}, \notag
\end{align}
$a$ and $a^{\dagger}$ satisfy $[a_{i}, a^{\dagger}_{j}]=\delta_{i,j}$.
We also define
\begin{align}
2(\ell_{+1})_j &= \{ (g_{+1/2})_j, (g_{+1/2})_j \} = 2( g_{+1/2})_j^2,
\label{eq:ell+12g} \\
2(\ell_{-1})_j &= \{ (g_{-1/2})_j, (g_{-1/2})_j \} = 2( g_{-1/2})_j^2,
\label{eq:ell-12g} \\
2(\ell_{0})_j &= \{ (g_{+1/2})_j, (g_{-1/2})_j \} = a_j^{\dagger}a_j + \frac{1}{2}.
\label{eq:ell02g}
\end{align}
As in $T^{2}$ case, $(g_{r})_j$ $(r=\pm 1/2)$ and $(\ell_{n})_j$ $(n=0,\pm1)$ in Eqs.~(\ref{eq:g-122g})-(\ref{eq:ell02g}) satisfy $OSp(1|2)_j$ algebra with $i=1,...,g$.
Note that $(\ell_{n})_j$ $(n=0,\pm1)$ satisfy $SL(2,\mathbb{R})_j$ subalgebra.
In particular, since the following relations,
\begin{align}
[2(\ell_{0})_j, (\ell_{\pm1})_j] = \mp 2 (\ell_{\pm1})_j, \quad [2(\ell_{0})_j, (g_{\pm1/2})_j] = \mp (g_{\pm1/2})_j,
\label{eq:ladder2g}
\end{align}
are satisfied, $(g_{-1/2})_j$ ($(g_{+1/2})_j$) and $(\ell_{-1})_j$ ($(\ell_{+1})_j$) raise (lower) the eigenvalue of $2(\ell_0)_j$ by $1$ and $2$, respectively.

Wave functions on the magnetized $T^{2g}$ with the unite charge $q=1$, $\Phi$, satisfy the following boundary conditions:
\begin{align}
\Phi(z+e_k) &= e^{2\pi i {^t}\alpha^S e_k } e^{\pi i[ {^t}N ({\rm Im}\Omega)^{-1} {\rm Im}z]_k} \Phi(z), \label{eq:BCek} \\
\Phi(z+\Omega e_k) &= e^{2\pi i {^t}\beta^S e_k} e^{\pi i[{\rm Im}\{(N\bar{\Omega}) ({\rm Im}\Omega)^{-1} z \}]_k} \Phi(z),
\label{eq:BCOmegaek}
\end{align}
where $\alpha^S = {^t}(\alpha^S_1,...,\alpha^S_g)$ and $\beta^S = {^t}(\beta^S_1,...,\beta^S_g)$ denote the Scherk-Schwarz phases.
Here, we consider a scalar, $\phi$, which satisfies the following equation of motion:
\begin{align}
-2[{^t}(\alpha^{\dagger}\alpha)^{-1}]_{\bar{j}i} \{ D_{z^{i}}, D_{\bar{z}^{\bar{j}}} \} \phi(z) &= m^2_{{\rm scalar}} \phi(z).
\label{eq:eomscalar0}
\end{align}
The left-hand side can be rewritten as
\begin{align}
({\rm LHS})
=&-2[{^t}(\alpha^{\dagger}\alpha)^{-1}]_{\bar{j}i} \{ D_{z^{i}}, D_{\bar{z}^{\bar{j}}} \} \phi(z) \notag \\
=& -2[{^t}(\alpha^{\dagger}\alpha)^{-1}]_{\bar{j}i} \left( 2 D_{z^{i}} D_{\bar{z}^{\bar{j}}} - \pi [{^t}N ({\rm Im}\Omega)^{-1}]_{i\bar{j}} \right) \phi(z) \notag \\
=& \left( -4 D_{z} \alpha^{-1} (\alpha^{\dagger})^{-1} {^t}D_{\bar{z}} + 2\pi {\rm tr}({^t}\alpha^{-1}{^t}N({\rm Im}\Omega)^{-1}{^t}(\alpha^{\dagger})^{-1}) \right) \phi(z) \notag \\
=& \left( 8\pi g_{-1/2} {\rm diag}\left(M |\alpha_0|^{-2} \right) {^t}g_{+1/2} 
+ 2\pi {\rm tr} {\rm diag}(M |\alpha_0|^{-2})  \right) \phi(z) \notag \\
=& \sum_{k=1}^{g} \frac{4\pi M_k}{|\alpha_0|_k^2} \left( 2(g_{-1/2})_k (g_{+1/2})_k + \frac{1}{2} \right) \phi(z) \notag \\
=& \sum_{k=1}^{g} \frac{4\pi M_k}{|\alpha_0|_k^2} \left( a_k^{\dagger} a_k + \frac{1}{2} \right) \phi(z) \notag \\
=& \sum_{k=1}^{g} \frac{4\pi M_k}{|\alpha_0|_k^2} 2(\ell_{0})_k \phi(z).
\label{eq:Laplacian}
\end{align}
The lowest state, $\phi=\Phi_{(0,...,0)}$, in particular, satisfies
\begin{align}
D_{\bar{z}^{j}} \Phi_{(0,...,0)}(z) = a_j \Phi_{(0,...,0)}(z) = (g_{+1/2})_j \Phi_{(0,...,0)}(z) = (\ell_{+1})_j \Phi_{(0,...,0)}(z) = 0,
\label{eq:lowest}
\end{align}
for $\forall j=1,...,g$.
Under the boundary conditions in Eqs.~(\ref{eq:BCek}) and (\ref{eq:BCOmegaek}), the solution of Eq.~(\ref{eq:lowest}) can be written as
\begin{align}
\Phi_{(0,...,0)}^{J}(z,\Omega,U)
=& [{\rm det}(U)]^{-1/2} [{\rm det}({\rm Im}\Omega)]^{1/4} \phi_{(0,...,0)}^{J}(z,\Omega) \notag \\
=& [{\rm det}(U)]^{-1/2} [{\rm det}({\rm Im}\Omega)]^{1/4} \left(\frac{{\rm det}(2N)}{V^2}\right)^{1/4} e^{2\pi i{^t}(J+\alpha^S)N^{-1}\beta^S} e^{\pi i {^t}z {^t}N ({\rm Im}\Omega)^{-1} {\rm Im}z} \notag \\
&\times \vartheta
\begin{bmatrix}
{^t}(J+\alpha^S)N^{-1}\\
-{^t}\beta
\end{bmatrix}
(Nz, N\Omega) \notag \\
=& [{\rm det}(U)]^{-1/2} [{\rm det}({\rm Im}\Omega)]^{1/4} \left(\frac{{\rm det}(2N)}{V^2}\right)^{1/4} e^{2\pi i{^t}(J+\alpha^S)N^{-1}\beta^S} e^{\pi i {^t}z {^t}N ({\rm Im}\Omega)^{-1} {\rm Im}z} \notag \\
&\times \sum_{\ell \in \mathbb{Z}^g}
e^{\pi i {^t}((J+\alpha^S)N^{-1}+\ell)N\Omega((J+\alpha^S)N^{-1}+\ell)}
e^{2\pi i {^t}((J+\alpha^S)N^{-1}+\ell)(Nz-\beta)},
\label{eq:wavlowest2g}
\end{align}
for $\forall J \in \Lambda_N$~\cite{Cremades:2004wa,Antoniadis:2009bg},
where $\Lambda_N$ denotes the lattice cell spanned by the lattice vectors,
\begin{align}
{^t}Ne_k \ (k=1,...,g).
\label{eq:latticevector}
\end{align}
Then, the excited state, $\phi=\Phi_{(n_1,...,n_g)}$, can be written by
\begin{align}
\Phi_{(n_1,...,n_g)}^{J}(z,\Omega,U) = \prod_{k=1}^{g} \left( \frac{1}{\sqrt{n_k !}} \left( \sqrt{2} (g_{-1/2})_k \right)^{n_k} \right) \Phi_{(0,...,0)}^{J}(z,\Omega,U).
\label{eq:excitedstate}
\end{align}
From Eqs.~(\ref{eq:eomscalar0}) and (\ref{eq:Laplacian}), the mass eigenvalue, $m^2_{{\rm scalar}}$, is given by
\begin{align}
m^2_{{\rm scalar}}
= \sum_{k=1}^{g} \frac{4\pi M_k}{|\alpha_0|_k^2} \left( n_k + \frac{1}{2} \right),
\label{eq:mass}
\end{align}
where $n_k + 1/2$ is the eigenvalue of $2(\ell_0)_k$.
The wave function, $\Phi_{(n_1,...,n_g)}^{J}$, is also eigenfunction of $2\ell_0$ defined by
\begin{align}
2\ell_0 \equiv \frac{1}{g}\sum_{k=1}^{g} 2(\ell_0)_k,
\label{eq:2ell02g}
\end{align}
with eigenvalue, $1/g\sum_{k=1}^{g}(n_k + 1/2)$,~i.e.,
\begin{align}
2\ell_0 \Phi_{(n_1,...,n_g)}^{J} = \left( \frac{1}{g}  \sum_{k=1}^{g}\left(n_k + \frac{1}{2}\right)\right) \Phi_{(n_1,...,n_g)}^{J}.
\label{eq:eigenfunc2ell0}
\end{align}

In the case of a spinor, $\psi$, it satisfies the following equation of motion:
\begin{align}
i \slashed{D} \psi(z) = m_{{\rm spinor}} \psi(z),
\label{eq:Diraceq}
\end{align}
where $i\slashed{D}$ denotes the Dirac operator and $(i\slashed{D})^2$ becomes the diagonalized operator.
Therefore, we can obtain the equation of motion for each spinor component.
First, the spinor component whose chirality on $z^j$  ($\forall j=1,...,g$) is positive, $\psi_{(+,...,+)}$, satisfies
\begin{align}
\left( \sum_{k=1}^{g} \frac{4\pi M_k}{|\alpha_0|_k^2}  2(g_{-1/2})_k(g_{+1/2})_k \right) \psi_{(+,...,+)}(z) = m_{{\rm spinor}}^2 \psi_{(+,...,+)}(z),
\label{eq:EoMall+}
\end{align}
and then we can obtain the solution:
\begin{align}
\psi_{(+,...,+)} = \Phi^{J}_{(n_1,...,n_g)},
\label{eq:solall+}
\end{align}
with
\begin{align}
m_{{\rm spinor}}^2 = \sum_{k=1}^{g} \frac{4\pi M_k}{|\alpha_0|_k^2} n_k.
\label{eq:massspinor}
\end{align}
Next, the spinor component whose chirality on $z^j$ ($\exists j=1,...,g$) is only negative, $\psi_{(+,...,+,-_j,+,...,+)}$, satisfies
\begin{align}
\left( \sum_{k=1}^{g} \left(\frac{4\pi M_k}{|\alpha_0|_k^2} 2(g_{-1/2})_k(g_{+1/2})_k \right) + \frac{4\pi M_j}{|\alpha_0|_j^2} \right) \psi_{(+,...,+)}(z) = m_{{\rm spinor}}^2 \psi_{(+,...,+)}(z),
\label{eq:EoMj-}
\end{align}
where we use Eqs.~(\ref{eq:DDbarF})-(\ref{eq:alpha0}) and $m_{{\rm spinor}}^2$ is in Eq.~(\ref{eq:massspinor}).
Then, the solution is given by
\begin{align}
\psi_{(+,...,+,-_j,+,...,+)} = \Phi_{(n_1,...n_{j-1},n_{j}-1,n_{j+1},...,n_{g})}.
\label{eq:solj-}
\end{align}
Namely, when the chirality on $z^j$ is negative, the level becomes $n_j -1$, where $n_j$ is the level in the case of the positive chirality on $z^j$.

In the next subsection, we extend the analysis of periodic functions of the Siegel modular forms in the previous subsection to those of the wave functions on a magnetized $T^{2g}$ in Eq.~(\ref{eq:excitedstate}).
Here, in order for the wave functions in Eq.~(\ref{eq:excitedstate}) to behave as the Siegel modular forms with the automorphy factor $j_h(g, \Omega_0)=[{\rm det}(C\Omega_0+D)]^h$, it is required that $n_1=\cdots=n_g \equiv n$.
In this case, the wave functions, $\Phi_{n} \equiv \Phi_{(n,...,n)}$ in Eq.~(\ref{eq:excitedstate}) can be written as
\begin{align}
\Phi_{n}^{J}(z,\Omega,U)
&= \frac{1}{(n!)^{g/2}} \left( \prod_{k=1}^{g} \sqrt{2} (g_{-1/2})_k \right)^n \Phi_{0}^{J}(z,\Omega,U) \notag \\
&= [{\rm det}(U)]^{(n+1/2)} [{\rm det}({\rm Im}\Omega)]^{(n+1/2)/2} \left(\frac{(-1)^{n}}{\sqrt{\pi^{n} n!}} \right)^{g} [{\rm det}(N)]^{-n/2} \left( \prod_{k=1}^{g} (D_z O)_k
\right)^n \phi_{0}^{J}(z,\Omega) \notag \\
&= [{\rm det}(U)]^{(n+1/2)} [{\rm det}({\rm Im}\Omega)]^{(n+1/2)/2} \phi_{n}^{J}(z,\Omega),
\label{eq:excitedstaten}
\end{align}
and they are eigenfunctions of $2\ell_0$ with the eigenvalue $n+1/2$.
We note that since $m_{\rm scalar}^2$ is written by
\begin{align}
m_{{\rm scalar}}^2 = \left( \sum_{k=1}^{g} \frac{4\pi M_k}{|\alpha_0|_k^2} \right) \left( n + \frac{1}{2} \right) \equiv \lambda \left( n + \frac{1}{2} \right),
\label{eq:massre}
\end{align}
the eigenvalue of $2\ell_0$ corresponds to the mass level.
In addition, when we define
\begin{align}
g_{\pm 1/2} = \prod_{k=1}^{g} (g_{\pm1/2})_k, \quad
\ell_{\pm 1} = \prod_{k=1}^{g} (\ell_{\pm1})_k,
\label{eq:detgell}
\end{align}
they satisfy the following algebra:
\begin{align}
[2\ell_{0}, \ell_{\pm1
}] = \mp2 \ell_{\pm1}, \quad [2\ell_0, g_{\pm 1/2}] \mp g_{\pm 1/2},
\label{eq:ladderdet2g}
\end{align}
which means that $g_{-1/2}$ ($g_{+1/2}$) and $\ell_{-1}$ ($\ell_{+1}$) raise (lower) the eigenvalue of $2\ell_0$ by $1$ and $2$, respectively.

Before the end of this subsection, we comment on a spinor case.
When the wave functions of the spinor component, $\psi_{(+,...,+)}$, (whose chirality on $z^j$ ($\forall j=1,...,g$) is positive,) behave as the Siegel modular forms with the automorphy factor $j_h(g, \Omega_0)=[{\rm det}(C\Omega_0+D)]^h$, only the component, $\psi_{(-,...,-)}$, (whose chirality on $z^j$ ($\forall j=1,...,g$) is negative,) also behave as the Siegel modular forms with the automorphy factor $j_h(g, \Omega_0)=[{\rm det}(C\Omega_0+D)]^h$, due to Eq.~(\ref{eq:solj-}) and $n_1=\cdots=n_g$.


\subsection{Periodic functions of wave functions}
\label{subsec:periodicwavefunctionT2g}

In this subsection, we extend the analysis of the periodic functions of Siegel modular forms in subsection~\ref{subsec:periodicSiegelmodularform} to those of wave functions.

As the analysis in Ref.~\cite{Kikuchi:2023awe},
$\phi^{J}_{n}$ in $\Phi^{J}_{n}$ in Eq.~(\ref{eq:excitedstaten}) (with the proper conditions of magnetic flux elements and Scherk-Schwarz phases) behave as Siegel modular forms of weight $1/2$,
where we take the normalization of $\phi^{J}_{n}$ as
\begin{align}
\int d^{g}zd^{g}\bar{z} \left( \phi^{J}_{n} \right)^{\ast} \phi^{K}_{n} = [{\rm det}({\rm Im}\Omega)]^{-(n+1/2)} \delta_{J,K}.
\label{eq:normalizationphiT2g}
\end{align}
As in Eq.~(\ref{eq:periodicSiegelmodular}), $\Phi^{J}_{n}$ become the periodic functions,~i.e.,
\begin{align}
\Phi_n^{J}(\gamma'(z, \Omega, U)) = \Phi_n^{J}(z, \Omega, U) \quad \gamma' \in G_d,
\label{eq:periodicwave2g}
\end{align}
and they are canonically normalized as
\begin{align}
\int d^{g}zd^{g}\bar{z} \left( \Phi^{J}_{n} \right)^{\ast} \Phi^{K}_{n} = \delta_{J,K}.
\label{eq:normalizationPhiT2g}
\end{align}
The lowest state, $\Phi^{J}_{0}$, also satisfies
\begin{align}
\hat{F}_j \Phi^{J}_{0}(z,\Omega,U) = 0,
\label{eq:Fhat2Phi02g}
\end{align}
where $\hat{F}_j$ is defined by modifying $F_j$ in Eq.~(\ref{eq:Fj2g}) as
\begin{align}
\hat{F}_j 
= e^{+2i\theta_j} \left( -2i[{^t}({\rm Im}z)O]_j [\partial_{\bar{z}}O]_j -2i{\rm Im}\tau_j \partial_{\bar{\tau}_j} + \frac{1}{2} e^{i\theta_j}\partial_{e^{i\theta_j}} \right).
\label{eq:Fhat2g}
\end{align}
Similarly, we define $\hat{E}_j$ by modifying $E_j$ in Eq.~(\ref{eq:Ej2g}) as
\begin{align}
\hat{E}_j 
= e^{-2i\theta_j} \left( 2i[{^t}({\rm Im}z)O]_j [\partial_{z}O]_j  + 2i{\rm Im}\tau_j \partial_{\bar{\tau}_j} - \frac{1}{2} e^{i\theta_j} \partial_{e^{i\theta_j}} \right).
\label{eq:Ehat2g}
\end{align}
They satisfy the following $SL(2,\mathbb{R})_j$ algebra:
\begin{align}
[\hat{E}_j, \hat{F}_j] = \hat{H}_j, \quad [\hat{H}_j, \hat{E}_j] = +2\hat{E}_j, \quad [\hat{H}_j, \hat{F}_j] = -2\hat{F}_j, \label{eq:SL2Ralghat2g}
\end{align}
where $\hat{H}_j$ is the same as $H_j$,
\begin{align}
\hat{H}_j = H_j = - e^{i\theta_j} \partial_{e^{i\theta_j}}.
\label{eq:Hhat2g}
\end{align}
In addition, $\Phi^{j}_{n}$ satisfies
\begin{align}
\hat{H}_j \Phi^{J}_{n}(z,\Omega,U) = \left( n + \frac{1}{2} \right) \Phi^{J}_{n}(z,\Omega,U).
\label{eq:Hhat2gPhi}
\end{align}
Here, any eigenvalues of $\hat{H}_j$ for $j=1,...,g$ are the same as the modular weight $h = n+1/2$, which is the eigenvalue of
\begin{align}
\hat{H} =& - {\rm det}U \partial_{{\rm det}U} = \frac{1}{g} \sum_{k=1}^{g} \hat{H}_j = H.
\label{eq:Hhatave}
\end{align}
We also define
\begin{align}
\hat{E} =& \prod_{k=1}^{g} \hat{E}_k, \label{eq:detEhat} \\
\hat{F} =& \prod_{k=1}^{g} \hat{F}_k, \label{eq:detFhat}
\end{align}
and they satisfy the following algebra:
\begin{align}
[\hat{H}, \hat{E}] = +2 \hat{E}, \quad [\hat{H}, \hat{F}] = -2 \hat{F},
\label{eq:HdetEFhatcond}
\end{align}
which means that $\hat{E}$ ($\hat{F}$) raises (lowers) the modular weight by $2$.

The modular weight $h=n+1/2$ is the same as the mass level of the scalar.
As for the spinor, on the other hand, we should modify the definition of the periodic functions of the wave functions.
In the following, we consider the scalar, $\phi$, and the spinor components, $\psi_{(+,,,.,+)} \equiv \psi_{+}$ and $\psi_{(-,...,-)} \equiv \psi_{-}$, which behave as the Siegel modular forms with the automorphy factors, $j_h(g, \Omega_0)=[{\rm det}(C\Omega_0+D)]^h$, $j_{h_{+}}(g, \Omega_0)=[{\rm det}(C\Omega_0+D)]^{h_+}$, and $j_{h_{-}}(g, \Omega_0)=[{\rm det}(C\Omega_0+D)]^{h_{-}}$, respectively.

First, recall that $z$ depends on $U$ and $U$ is written as in Eq.~(\ref{eq:U}).
Here, we introduce
\begin{align}
U = O {\rm diag}(e^{i\theta'}) {O'}^{-1},
\label{eq:Uprime}
\end{align}
where ${\theta'}_j=\theta_j+\Delta \theta_j$.
Then, $z'$ is related to $z$ as
\begin{align}
O^{-1} z' = {\rm diag}( ({\rm Im}\tau)^{1/2} e^{-i{\theta'}} ) {O'}^{-1} z_0 = {\rm diag}(e^{-i\Delta \theta}) {\rm diag}( ({\rm Im}\tau)^{1/2} e^{-i\theta} ) {O'}^{-1} z_0 = {\rm diag}(e^{-i\Delta \theta}) O^{-1} z.
\label{zprimez2g}
\end{align}
Now, let us reconsider the definition of periodic functions of wave functions.
In the case of a scalar $\phi$, the periodic function, $\Phi$, with the modular weight $h$ satisfies that
\begin{align}
\Phi(O^{-1} z',\Omega,U')
=& [{\rm det}(U')]^{-h} [{\rm det}({\rm Im}\Omega)]^{h/2} \phi( O^{-1} z',\Omega) \notag \\
=& e^{-ih\sum_{j}\Delta\theta_j} [{\rm det}(U)]^{-h} [{\rm det}({\rm Im}\Omega)]^{h/2} \phi({\rm diag}(e^{-i\Delta\theta}) O^{-1} z,\Omega) \notag \\
=& e^{-ih\sum_{j}\Delta\theta_j} \left( [{\rm det}(U)]^{-h} [{\rm det}({\rm Im}\Omega)]^{h/2} \phi( O^{-1} z,\Omega) \right) \notag \\
=& e^{-ih\sum_{j}\Delta\theta_j} \Phi(O^{-1} z,\tau,\theta),
\label{eq:rePhi2g}
\end{align}
where the scalar satisfies that
\begin{align}
\phi({\rm diag}(e^{-i\Delta\theta}) O^{-1} z,\Omega) = \phi( O^{-1} z,\Omega).
\label{eq:scalar2g}
\end{align}
In the case of a spinor $\psi_{\pm}$, on the other hand, the periodic function, $\Psi_{\pm}$ with the modular weight $h_{\pm}$ satisfies that
\begin{align}
\Psi_{\pm}(O^{-1} z',\Omega,U')
=& [{\rm det}(U')]^{-h_{\pm}} [{\rm det}({\rm Im}\Omega)]^{h_{\pm}/2} \psi_{\pm}(O^{-1} z',\Omega) \notag \\
=& e^{-ih_{\pm}\sum_{j}\Delta\theta_j} [{\rm det}(U)]^{-h_{\pm}} [{\rm det}({\rm Im}\Omega)]^{h_{\pm}/2} \psi_{\pm}({\rm diag}(e^{-i\Delta\theta}) O^{-1} z,\Omega) \notag \\
=& e^{-ih_{\pm}\sum_{j}\Delta \theta_j} e^{\mp i\sum_{j}\Delta \theta_j/2} \left( [{\rm det}(U)]^{-h_{\pm}} [{\rm det}({\rm Im}\Omega)]^{h_{\pm}/2} \psi_{\pm}(O^{-1} z,\Omega) \right) \notag \\
=& e^{-i\left( h_{\pm} \pm 1/2 \right) \sum_{j} \Delta \theta_j} \Psi_{\pm}(O^{-1} z,\Omega, U),
\label{eq:rePsi2g}
\end{align}
where the spinor satisfies that
\begin{align}
\psi_{\pm}({\rm diag}(e^{-i\Delta\theta}) O^{-1} z,\Omega) = e^{\mp i\sum_{j}\Delta \theta_j/2} \psi_{\pm}(O^{-1} z,\Omega).
\label{eq:spinor2g}
\end{align}
From Eqs.~(\ref{eq:rePhi2g}) and (\ref{eq:rePsi2g}),
when the periodic function of the wave function of the spinor, $\Psi_{\pm}$, is the same as the wave function of the scalar, $\Phi$, the modular weight of the spinor, $h_{\pm}$, can be written by the modular weight of the scalar, $h$, as $h_{\pm}=h \mp 1/2$.
In this case, the spinor, $\psi_{\pm}$, in Eq.~(\ref{eq:rePsi2g}), which satisfies Eq.~(\ref{eq:scalar2g}), can be represented by the scalar, $\phi$, in Eq.~(\ref{eq:rePhi2g}), which satisfies Eq.~(\ref{eq:spinor2g}), as
\begin{align}
\psi_{\pm}(z,\Omega) = e^{\mp i\sum_{j}\theta_j/2} \phi(z,\tau).
\label{eq:spinorscalar2g}
\end{align}
Hence, the modular weight of the spinor, $\psi_{+,n}=\Phi_{n}$ and $\psi_{-,n}=\Phi_{n-1}$, can be obtained as
\begin{align}
\hat{H} \psi_{+,n} =& \left( \left( n+\frac{1}{2} \right) - \frac{1}{2} \right) \psi_{+,n} = n \psi_{+,n} = h_{+} \psi_{+,n},
\label{eq:weightpsi+n} \\
\hat{H} \psi_{-,n} =& \left( \left( (n-1)+\frac{1}{2} \right) + \frac{1}{2} \right) \psi_{-,n} = n \psi_{+,n} = h_{-} \psi_{-,n}.
\label{eq:weightpsi-n}
\end{align}
Note that it is also the same as the mass level of the spinor in Eq.~(\ref{eq:massspinor}) with $n_k=n$ ($\forall k=1,...,g$).


\subsection{Modular weight versus Mass level}
\label{subsec:modularweightmasslevel2g}

In this subsection, we show that the modular weight of the wave functions on a magnetized $T^{2g}$ is equivalent to their mass level.
Here, we consider the scalar, $\phi=\Phi_{n}$.

From Eqs.~(\ref{eq:Hhat2gPhi}) and (\ref{eq:eigenfunc2ell0}) with $n_1=\cdots=n_g\equiv n$, both the modular weight (which is the eigenvalue of $\hat{H}$) and the mass level (which is the eigenvalue of $2\ell_{0,{\rm scalar}}$) are $n+1/2$.
From Eqs.~(\ref{eq:HdetEFhatcond}) and (\ref{eq:ladderdet2g}), the modular weight and the mass level are raised (lowered) by 2 by operating $\hat{E}$ ($\hat{F}$) and $\ell_{-1}$ ($\ell_{+1}$), respectively.
In particular, from Eqs.~(\ref{eq:Fhat2Phi02g}) and (\ref{eq:lowest}), $\Phi_{0}$ becomes the lowest state with respect to both the modular weight and the mass level.
Now, as shown in Eq.~(\ref{eq:ellEhat2gdetail}) of Appendix~\ref{apsec:ellEhat2g} in detail,
we can show that operating $\ell_{-1}$ on $\Phi_{0}$ is equivalent to operating $\hat{E}$ on $\Phi_{0}$;
\begin{align}
\ell_{-1} \Phi^j_0(z,\Omega,U) = \hat{E} \Phi^J_0(z,\Omega,U).
\label{eq:ellEhat2g}
\end{align}
Similarly, operating $\ell_{+1}$ is equivalent to $-\hat{F}$.\footnote{Indeed, it is consistent that $[(\ell_{+1})_j, (\ell_{-1})_j]=2(\ell_{0})_j$ and $[-\hat{F}_j, \hat{E}_j]=\hat{H}_j$.}
In other words, $\ell_{-1}$ ($\ell_{+1}$) raises (lowers) not only the mass level but the modular weight by 2.
Furthermore, $g_{-1/2}$ ($g_{+1/2}$) also raises (lowers) not only the mass level but the modular weight by 1.
Indeed, they satisfy the following algebra:
\begin{align}
[\hat{H}, \ell_{\pm1}] = \mp 2 \ell_{\pm1}, \quad [\hat{H}, g_{\pm1/2}] = \mp g_{\pm1/2}.
\label{eq:Hhat2gellg}
\end{align}
Therefore, it is found that the modular weight of the wave functions on magnetized $T^{2g}$ is equivalent to their mass level.


\section{Conclusion}
\label{sec:conclusion}

In previous studies~\cite{Kikuchi:2020frp,Kikuchi:2021ogn}, wave functions of the lowest states on magnetized $T^{2}$ behave as modular forms of weight $1/2$.
In Ref.~\cite{Kikuchi:2023clx}, it has been suggested that the modular weight is equivalent to the mass level.
In this paper, we have shown it explicitly in section~\ref{sec:T2}; we have shown that the modular weight equals to the mass level and also ladder operators related to the modular weight, which are operators extended from ones in Ref.~\cite{Ding:2020zxw}, are equivalent to ladder operators related to the mass level.

We also extended the analysis to magnetized $T^{2g}$ models in section~\ref{sec:T2g}.
In order to do it, we extended the analysis in section~4 in Ref.~\cite{Ding:2020zxw} in detail, and we constructed wave functions of the excited states on magnetized $T^{2g}$.
Similarly to the magnetized $T^{2}$ case, we have found that the modular weight is equivalent to the mass level.

Magnetized D-brane models and intersecting D-brane models are T-dual to each other \cite{Ibanez:2012zz}. In the intersecting D-brane side, low energy effective field theory is analyzed by use of conformal field theory \cite{Cvetic:2003ch,Abel:2003vv}. It is important to compare our results with them. We would study this issue elsewhere.


\section*{Acknowledgments}

This work was supported by JSPS KAKENHI Grant Numbers JP23K03375 (T.K.), JP25H01539 (H.O.), 
and JST SPRING, Grant Number JPMJSP2119 (T.J.), and the YAMAGUCHI UNIVERSITY FUND and the Sumitomo Foundation under Grant for Basic Science Research (Grant No. 2502479)(M.T.).


\appendix

\section{Equivalence of $\ell_{-1}$ and $\hat{E}$ in magnetized $T^2$ model}
\label{apsec:ellEhat}

Here, we show the detailed calculation of Eq.~(\ref{eq:ellEhat}) in the following.
\begin{align}
&\ell_{-1} \Phi^j_0(z,\tau,\theta) \notag \\
=& e^{-2i\theta} ({\rm Im}\tau) \frac{1}{2\pi M} (\partial_{z} - \frac{\pi M}{2{\rm Im}\tau}\bar{z})^2 e^{-i\theta/2} ({\rm Im}\tau)^{1/4} \phi^{j}_{0}(z,\tau) \notag \\
=& e^{-5i\theta/2} ({\rm Im}\tau)^{5/4} \frac{1}{2\pi M} (e^{\pi iM\bar{z}\frac{{\rm Im}z}{{\rm Im}\tau}} \partial_{z}^{2} e^{-\pi iM\bar{z}\frac{{\rm Im}z}{{\rm Im}\tau}}) \left(\frac{2M}{A} \right)^{1/4} e^{2\pi i\frac{j+\alpha^S}{M}\beta^S} 
e^{\pi iMz\frac{{\rm Im}z}{{\rm Im}\tau}} 
\vartheta
\begin{bmatrix}
\frac{j+\alpha^S}{M} \\ -\beta^S
\end{bmatrix}
(Mz,M\tau) \notag \\
=& e^{-5i\theta/2} ({\rm Im}\tau)^{5/4} \frac{1}{2\pi M} \left(\frac{2M}{A} \right)^{1/4} e^{2\pi i\frac{j+\alpha^S}{M}\beta^S} e^{\pi iMz\frac{{\rm Im}z}{{\rm Im}\tau}} \notag \\
&\times e^{2\pi M\frac{({\rm Im}z)^2}{{\rm Im}\tau}} \partial_{z}^{2} e^{-2\pi M\frac{({\rm Im}z)^2}{{\rm Im}\tau}} \sum_{\ell \in \mathbb{Z}} e^{\pi iM\tau (\frac{j+\alpha^S}{M}+\ell)^2} e^{2\pi i(Mz-\beta^S)(\frac{j+\alpha^S}{M}+\ell)} \notag \\
=& e^{-5i\theta/2} ({\rm Im}\tau)^{5/4} \frac{1}{2\pi M} \left(\frac{2M}{A} \right)^{1/4} e^{2\pi i\frac{j+\alpha^S}{M}\beta^S} e^{\pi iMz\frac{{\rm Im}z}{{\rm Im}\tau}} \notag \\
&\times (\partial_{z} + 2\pi iM\frac{{\rm Im}z}{{\rm Im}\tau}) (\partial_{z} + 2\pi iM\frac{{\rm Im}z}{{\rm Im}\tau}) \sum_{\ell \in \mathbb{Z}} e^{\pi iM\tau (\frac{j+\alpha^S}{M}+\ell)^2} e^{2\pi i(Mz-\beta^S)(\frac{j+\alpha^S}{M}+\ell)} \notag \\
=& e^{-5i\theta/2} ({\rm Im}\tau)^{4/5} \frac{1}{2\pi M} \left(\frac{2M}{A} \right)^{1/4} e^{2\pi i\frac{j+\alpha^S}{M}\beta^S} e^{\pi iMz\frac{{\rm Im}z}{{\rm Im}\tau}} \notag \\
&\times \left(\partial_{z}^2 + 4\pi iM\frac{{\rm Im}z}{{\rm Im}\tau}\partial_{z} + \frac{\pi M}{{\rm Im}\tau} + \left( 2\pi iM \frac{{\rm Im}z}{{\rm Im}\tau} \right)^2 \right) \sum_{\ell \in \mathbb{Z}} e^{\pi iM\tau (\frac{j+\alpha^S}{M}+\ell)^2} e^{2\pi i(Mz-\beta^S)(\frac{j+\alpha^S}{M}+\ell)} \notag \\
=& e^{-5i\theta/2} ({\rm Im}\tau)^{4/5} \frac{1}{2\pi M} \left(\frac{2M}{A} \right)^{1/4} e^{2\pi i\frac{j+\alpha^S}{M}\beta^S} e^{\pi iMz\frac{{\rm Im}z}{{\rm Im}\tau}} \notag \\
&\times \left(4\pi iM \partial_{\tau} + 4\pi iM\frac{{\rm Im}z}{{\rm Im}\tau}\partial_{z} + \frac{\pi M}{{\rm Im}\tau} + \left( 2\pi iM \frac{{\rm Im}z}{{\rm Im}\tau} \right)^2 \right) \sum_{\ell \in \mathbb{Z}} e^{\pi iM\tau (\frac{j+\alpha^S}{M}+\ell)^2} e^{2\pi i(Mz-\beta^S)(\frac{j+\alpha^S}{M}+\ell)} \notag \\
=& e^{-5i\theta/2} ({\rm Im}\tau)^{1/4} \left(\frac{2M}{A} \right)^{1/4} e^{2\pi i\frac{j+\alpha^S}{M}\beta^S} e^{\pi iMz\frac{{\rm Im}z}{{\rm Im}\tau}} \notag \\
&\times \left(2i {\rm Im}\tau \partial_{\tau} + 2i{\rm Im}z \partial_{z} + \frac{1}{2} - 2\pi M \frac{({\rm Im}z)^2}{{\rm Im}\tau} \right) \sum_{\ell \in \mathbb{Z}} e^{\pi iM\tau (\frac{j+\alpha^S}{M}+\ell)^2} e^{2\pi i(Mz-\beta^S)(\frac{j+\alpha^S}{M}+\ell)} \notag \\
=& e^{-5i\theta/2} ({\rm Im}\tau)^{1/4} \left(2i {\rm Im}\tau \partial_{\tau} + 2i{\rm Im}z \partial_{z} + \frac{1}{2} \right) \left(\frac{2M}{A} \right)^{1/4} e^{2\pi i\frac{j+\alpha^S}{M}\beta^S}  e^{\pi iMz\frac{{\rm Im}z}{{\rm Im}\tau}} 
\vartheta
\begin{bmatrix}
\frac{j+\alpha^S}{M} \\ -\beta^S
\end{bmatrix}
(Mz,M\tau) \notag \\
=& e^{-2i\theta} \left(2i {\rm Im}\tau \partial_{\tau} + 2i{\rm Im}z \partial_{z} + \frac{i}{2}\partial_{\theta} \right) e^{-i\theta/2} ({\rm Im}\tau)^{1/4} 
\phi^{j}_{0}(z,\tau) \notag \\
=& \hat{E} \Phi^j_0(z,\tau,\theta).
\label{eq:ellEhatdetail}
\end{align}


\section{Equivalence of $\ell_{-1}$ and $\hat{E}$ in magnetized $T^{2g}$ model}
\label{apsec:ellEhat2g}

Here, we show the detailed calculation of Eq.~(\ref{eq:ellEhat2g}) in the following.
\begin{align}
&\ell_{-1} \Phi^J_{0}(z,\Omega,U) \notag \\
=& \prod_{k=1}^{g} \left( \left[ \left(\partial_{z}-\frac{\pi}{2}[{^t}\bar{z}{^t}N({\rm Im}\Omega)^{-1}] \right) O \right]_k^2 \left( (2\pi M_k)^{-1} {\rm Im}\tau_k e^{-2i\theta_k} \right) \right) [{\rm det}(U)]^{-1/2} [{\rm det}({\rm Im}\Omega)]^{1/4} \phi^{J}_{0}(z,\Omega) \notag \\
=& [{\rm det}(U)]^{-5/2} [{\rm det}({\rm Im}\Omega)]^{5/4} [{\rm det}(2\pi N)]^{-1} e^{\pi i {^t}\bar{z} {^t}N ({\rm Im}\Omega)^{-1} {\rm Im}z}\prod_{k=1}^{g} \left[ \partial_{z} O \right]_k^2 e^{-\pi i {^t}\bar{z} {^t}N ({\rm Im}\Omega)^{-1} {\rm Im}z} \notag \\
&\times \left(\frac{{\rm det}(2N)}{V^2}\right)^{1/4} e^{2\pi i{^t}(J+\alpha^S)N^{-1}\beta^S} e^{\pi i {^t}z {^t}N ({\rm Im}\Omega)^{-1} {\rm Im}z} \vartheta
\begin{bmatrix}
{^t}(J+\alpha^S)N^{-1}\\
-{^t}\beta
\end{bmatrix}
(Nz, N\Omega) \notag \\
=& [{\rm det}(U)]^{-5/2} [{\rm det}({\rm Im}\Omega)]^{5/4} [{\rm det}(2\pi N)]^{-1} \left(\frac{{\rm det}(2N)}{V^2}\right)^{1/4} e^{2\pi i{^t}(J+\alpha^S)N^{-1}\beta^S} e^{\pi i {^t}z {^t}N ({\rm Im}\Omega)^{-1} {\rm Im}z} \notag \\
&\times e^{2\pi {^t}({\rm Im}z) {^t}N ({\rm Im}\Omega)^{-1} {\rm Im}z} \prod_{k=1}^{g} \left[ \partial_{z} O \right]_k^2 e^{-2\pi {^t}({\rm Im}z) {^t}N ({\rm Im}\Omega)^{-1} {\rm Im}z} \vartheta
\begin{bmatrix}
{^t}(J+\alpha^S)N^{-1}\\
-{^t}\beta
\end{bmatrix}
(Nz, N\Omega) \notag \\
=& [{\rm det}(U)]^{-5/2} [{\rm det}({\rm Im}\Omega)]^{5/4} [{\rm det}(2\pi N)]^{-1} \left(\frac{{\rm det}(2N)}{V^2}\right)^{1/4} e^{2\pi i{^t}(J+\alpha^S)N^{-1}\beta^S} e^{\pi i {^t}z {^t}N ({\rm Im}\Omega)^{-1} {\rm Im}z} \notag \\
&\times \left[ \left( \partial_{z} + 2\pi i {^t}({\rm Im}z){^t}N ({\rm Im}\Omega)^{-1} \right) O \right]_k^2 \vartheta
\begin{bmatrix}
{^t}(J+\alpha^S)N^{-1}\\
-{^t}\beta
\end{bmatrix}
(Nz, N\Omega) \notag \\
=& [{\rm det}(U)]^{-5/2} [{\rm det}({\rm Im}\Omega)]^{5/4} [{\rm det}(2\pi N)]^{-1} \left(\frac{{\rm det}(2N)}{V^2}\right)^{1/4} e^{2\pi i{^t}(J+\alpha^S)N^{-1}\beta^S} e^{\pi i {^t}z {^t}N ({\rm Im}\Omega)^{-1} {\rm Im}z} \notag \\
&\times \prod_{k=1}^{g} \left( \left[ \partial_{z}O \right]_k^2 + \left[ 4\pi i {^t}({\rm Im}z){^t}N ({\rm Im}\Omega)^{-1} O \right]_k \left[ \partial_{z}O \right]_k + \pi M_k ({\rm Im}\tau_k)^{-1} + \left[ 2\pi i {^t}({\rm Im}z){^t}N ({\rm Im}\Omega)^{-1} O \right]_k^2 \right) \notag \\
&\times \sum_{\ell \in \mathbb{Z}^g}
e^{\pi i {^t}((J+\alpha^S)N^{-1}+\ell)N\Omega((J+\alpha^S)N^{-1}+\ell)}
e^{2\pi i {^t}((J+\alpha^S)N^{-1}+\ell)(Nz-\beta)} \notag \\
=& [{\rm det}(U)]^{-5/2} [{\rm det}({\rm Im}\Omega)]^{5/4} [{\rm det}(2\pi N)]^{-1} \left(\frac{{\rm det}(2N)}{V^2}\right)^{1/4} e^{2\pi i{^t}(J+\alpha^S)N^{-1}\beta^S} e^{\pi i {^t}z {^t}N ({\rm Im}\Omega)^{-1} {\rm Im}z} \notag \\
&\times \prod_{k=1}^{g} \left( 4\pi i \partial_{\tau_k} + \left[ 4\pi i {^t}({\rm Im}z){^t}N ({\rm Im}\Omega)^{-1} O \right]_k \left[ \partial_{z}O \right]_k + \pi M_k ({\rm Im}\tau_k)^{-1} + \left[ 2\pi i {^t}({\rm Im}z){^t}N ({\rm Im}\Omega)^{-1} O \right]_k^2 \right) \notag \\
&\times \sum_{\ell \in \mathbb{Z}^g}
e^{\pi i {^t}((J+\alpha^S)N^{-1}+\ell)N\Omega((J+\alpha^S)N^{-1}+\ell)}
e^{2\pi i {^t}((J+\alpha^S)N^{-1}+\ell)(Nz-\beta)} \notag
\end{align}
\begin{align}
=& [{\rm det}(U)]^{-5/2} [{\rm det}({\rm Im}\Omega)]^{1/4} \left(\frac{{\rm det}(2N)}{V^2}\right)^{1/4} e^{2\pi i{^t}(J+\alpha^S)N^{-1}\beta^S} e^{\pi i {^t}z {^t}N ({\rm Im}\Omega)^{-1} {\rm Im}z} \notag \\
&\times \prod_{k=1}^{g} \left( 2 i {\rm Im}\tau_k \partial_{\tau_k} + 2i \left[ {^t}({\rm Im}z)O \right]_k \left[ \partial_{z}O \right]_k + \frac{1}{2} - 2\pi \left[  {^t}({\rm Im}z)O \right]_k^2 M_k ({\rm Im}\tau_k)^{-1} \right) \notag \\
&\times \vartheta
\begin{bmatrix}
{^t}(J+\alpha^S)N^{-1}\\
-{^t}\beta
\end{bmatrix}
(Nz, N\Omega) \notag \\
=& [{\rm det}(U)]^{-5/2} [{\rm det}({\rm Im}\Omega)]^{1/4} \prod_{k=1}^{g} \left( 2 i {\rm Im}\tau_k \partial_{\tau_k} + 2i \left[ {^t}({\rm Im}z)O \right]_k \left[ \partial_{z}O \right]_k + \frac{1}{2} \right) \notag \\
&\times \left(\frac{{\rm det}(2N)}{V^2}\right)^{1/4} e^{2\pi i{^t}(J+\alpha^S)N^{-1}\beta^S} e^{\pi i {^t}z {^t}N ({\rm Im}\Omega)^{-1} {\rm Im}z} \vartheta
\begin{bmatrix}
{^t}(J+\alpha^S)N^{-1}\\
-{^t}\beta
\end{bmatrix}
(Nz, N\Omega) \notag \\ 
=& [{\rm det}(U)]^{-5/2} [{\rm det}({\rm Im}\Omega)]^{1/4} \prod_{k=1}^{g} e^{-2i\theta_k} \left( 2 i {\rm Im}\tau_k \partial_{\tau_k} + 2i \left[ {^t}({\rm Im}z)O \right]_k \left[ \partial_{z}O \right]_k - \frac{1}{2} e^{i\theta_j} \partial_{e^{i\theta_j}} \right) \notag \\
&\times [{\rm det}(U)]^{-1/2} [{\rm det}({\rm Im}\Omega)]^{1/4} \phi^{J}_{0}(z,\Omega) \notag \\ 
=&\hat{E} \Phi^J_0(z,\Omega,U),
\label{eq:ellEhat2gdetail}
\end{align}


\section{Two-dimensional conformal field theory}
\label{apsec:CFT}

Here we give a brief review of two-dimensional classical conformal field theory to compare with modular forms and wave functions.
The complex coordinate $w$ transforms as  
\begin{align}
    w \to \frac{aw+b}{cw+d},
\end{align}
under the global conformal symmetry $SL(2,\mathbb{C}) \simeq SO(2,2)$,
where
\begin{align}
    SL(2,\mathbb{C}) \simeq SO(2,2) = \left\{ 
   \left. \begin{pmatrix}
     a &b  \\ c & d  
    \end{pmatrix} ~\right|~ a,b,c,d \in \mathbb{C}, ad-bc=1
    \right\}.
\end{align}
Their generators are written by 
\begin{align}
    L_w=-w^{n+1}\frac{\partial}{\partial w},
\end{align}
and they satisfy the following algebraic relations:
\begin{align}
    [L_m,L_n]=(m-n)L_{m+n},
\end{align}
with $m,n=-1,0,1$ for the global symmetry.

The chiral primary field $\Phi(z)$ with the conformal dimension d transforms as 
\begin{align}
    \varphi(w)=\varphi(w')\left(\frac{dw'}{dw}\right)^d,
\end{align}
under the conformal transformation, $w \to w'$.
Under the rotation,
\begin{align}
    w \to w'=e^{i\theta d}w,
\end{align}
the above field behaves as 
\begin{align}
    \varphi(w)=\varphi(w')e^{i\theta d}.
    \label{eq:CFT}
\end{align}

\bibliography{ref}{}
\bibliographystyle{JHEP} 

\end{document}